\documentclass[a4paper,11pt]{article}
\usepackage{jheppub}
\usepackage{graphicx}
\usepackage{amsmath}
\usepackage{amssymb}
\usepackage{bbold}
\usepackage{float}
\usepackage{slashed}
\usepackage{dcolumn}
\usepackage{bm}
\usepackage{color}
\usepackage{multirow}
\allowdisplaybreaks

\newcommand{\nn}{\nonumber}
\newcommand{\beq} {\begin{equation}}
\newcommand{\eeq} {\end{equation}}
\newcommand{\beqa} {\begin{eqnarray}}
\newcommand{\eeqa} {\end{eqnarray}}

\newcommand{\eps}{\epsilon}

\newcommand{\morder}[1]{{\cal O}\left(#1 \right)}

\def\m{\mu}
\def\n{\nu}
\def\r{\rho}
\def\s{\sigma}

\renewcommand{\Re}{\mathrm{Re}}

\newcommand{\uv}{\mathrm{UV}}
\newcommand{\ir}{\mathrm{IR}}

\newcommand{\be}{\begin{equation}}
\newcommand{\ee}{\end{equation}}
\newcommand{\bea}{\begin{eqnarray}}
\newcommand{\eea}{\end{eqnarray}}

\renewcommand{\l}{\lambda}

\newcommand{\dis}[1]{\begin{equation}\begin{split}#1\end{split}\end{equation}}

\preprint{
\begin{flushright}
APCTP Pre2025 - 010\\  
PNUTP-25/A01\\
CTPU-PTC-25-13
\end{flushright}
}

\title{Holographic analysis of near-conformal dynamics and light dilaton}

\affiliation[a]{Asia Pacific Center for Theoretical Physics,   Pohang 37673,   Korea }
\affiliation[b]{Department of Physics,  Pusan National University, Busan 46241,   Korea }
\affiliation[c]{Extreme Physics Institute, Pusan National University, Busan 46241, Korea}
\affiliation[d]{
Particle Theory and Cosmology Group, Center for Theoretical Physics of the Universe, \\
 Institute for Basic Science (IBS), Daejeon 34126, Korea \
}
\affiliation[e]{Department of Physics, Pohang University of Science and Technology,   Pohang 37673,   Korea }
\affiliation[f]{Departamento de F\'isica de Altas Energ\'ias, Instituto de Ciencias Nucleares, Universidad Nacional Aut\'onoma de M\'exico, Apartado Postal 70-543, CDMX 04510, M\'exico}

\author[a,f]{Jes\'us Cruz Rojas,}
\author[b,c]{Deog Ki Hong,}
\author[d]{Sang Hui Im,}
\author[a,e]{Matti J\"arvinen}

\emailAdd{jesus.cruz@correo.nucleares.unam.mx}
\emailAdd{dkhong@pusan.ac.kr}
\emailAdd{imsanghui@ibs.re.kr}
\emailAdd{matti.jarvinen@apctp.org}

\abstract{We carry out a detailed
analysis of the region slightly outside the conformal window of a non-trivial infrared  
fixed point in a generic bottom-up holographic setup.  We focus on models, which study the dynamics of a scalar field, dual to quark degrees of freedom, in a (nearly) AdS geometry. Such models realize the picture expected for vector-like near-conformal theories from Dyson-Schwinger analysis. The analysis covers a toy model, which allows for analytic solutions, and a more general setup as well, which encompasses a complete model for the ultraviolet physics.  We analyze the conditions for the appearance of a parametrically light scalar state in the spectrum, which can act as a candidate for the Pseudo-Nambu-Goldstone boson arising from the breaking of the approximate conformal symmetry. We also present detailed results for the vacuum structure, correlators, and Ward identities in the near-conformal regime.
}

\begin{document}
\maketitle

\section{Introduction}

Since the discovery of a light Higgs
boson at the LHC in 2012 \cite{ATLAS:2012yve, CMS:2012qbp}, there have been tremendous efforts to find a hint of new physics beyond the standard model (BSM). One of the expected features for new physics that are consistent with the current status of the experimental endeavors, indicating large separation of scales in BSM, is near conformality, characterized by a parametrically light bound state, corresponding to a pseudo-Nambu-Goldstone boson (PNGB) of the scale symmetry, spontaneously broken by the hyper-quark condensate.

The approach of explaining the extremely fine-tuned Higgs mass from the conformal edge, at which the BSM physics enjoys scale invariance at high energies, turns out to be quite attractive. One reason for this is because it nicely explains why there have not been any hint of new particles around the weak scale, $\Lambda_{ew}\sim 1$ TeV,  which has been vigorously explored at current energy frontiers. Another notable characteristic is that if an infrared scale is dynamically generated from the
quantum criticality, the modulus of the scale-invariant theory, the so-called dilaton, naturally develops a mass as well. If the dilaton is sufficiently light and weakly interacting, it can be a good candidate for dark matter \cite{Choi:2012kx,Hong:2017smd}.

An appealing approach for our study is to analyze the near-conformal physics from a strongly coupled theory that can be modeled by using the gauge/gravity duality. There have been several studies of holographic models on gauge theories that exhibit a conformal window, with some of them reporting a parametrically light mode, whereas other examples mention the lack of this. 
We focus here on models that study the dynamics of a scalar field, dual to quark degrees of freedom, in a (nearly) AdS geometry (see, e.g.~\cite{Matsuzaki:2012xx,Arean:2012mq,Evans:2013vca,Erdmenger:2020lvq, Erdmenger:2020flu, Pomarol:2019aae,CruzRojas:2023jhw}). Such models realize the picture expected for QCD-like theories from the Dyson-Schwinger analysis~\cite{Kaplan:2009kr}. For other holographic approaches discussing near-conformality and light modes, see, for example,~\cite{Nunez:2008wi,Elander:2010wd,Kutasov:2011fr,Anguelova:2012mu,Megias:2014iwa,Cox:2014zea}.   

In \cite{Matsuzaki:2012xx} the authors study a bottom-up holographic model, dual to a walking technicolor QFT. Here, they found a limit where the techni-dilaton becomes a massless Nambu-Goldstone boson for the broken scale symmetry and which naturally realizes a light techni-dilaton.
On the other hand in \cite{Arean:2012mq} the authors focus on the bottom-up V-QCD model, combining a glue sector, which is a model for large-N Yang Mills in four dimensions, and a flavour sector inspired by tachyon condensation in string theory. However, while in their study scalar mesons are among the lightest states, there is no sign of a parametrically light scalar state in the spectrum. 

There have also been other approaches like in \cite{Evans:2013vca, Erdmenger:2020lvq, Erdmenger:2020flu} where the authors present a simplified holographic model based on the D3-probe-D7 set-up for chiral symmetry breaking in gauge theory. Here, it is shown in an explicit example that if the gradient of the running of the anomalous dimension falls to zero at the scale where the Breitenlohner-Freedman (BF) bound violation occurs, the theory possesses a techni-dilaton state that is parametrically lighter than the dynamically generated quark mass. In \cite{Pomarol:2019aae,Pomarol:2023xcc} the authors provide another holographic model dual to a strongly-coupled theory close to the conformal to non-conformal transition, and find that the dilaton mode is always the lightest resonance, although not parametrically lighter than the others.

More recently in a previous work \cite{CruzRojas:2023jhw} we have also studied a simple holographic model for gauge theories near the conformal edge. By explicitly analyzing the
spectra of the composite states, choosing the parameters of the model such that the back reaction to the bulk geometry is negligible, we found that there is a unique scalar state whose mass is parametrically lighter than all other states. The lightest scalar is identified as the Nambu-Goldstone boson associated with the scale symmetry, by showing that it saturates at low energies the anomalous Ward identity of dilatation currents. 
There are also examples of holographic top-down models that contain light scalar modes, albeit using rather complex Klebanov--Strassler backgrounds~\cite{Elander:2017cle,Elander:2017hyr}. Simpler models also show light states, but in phases that are unstable (see, e.g.,~\cite{Elander:2021wkc}). Recently, a top-down model with a light dilaton state at the critical end-point of a line of first-order phase transition has also been constructed~\cite{Elander:2025fpk}.

In this article, we focus on studying gauge
theories that exhibit dynamics slightly outside the conformal windows of non-trivial infrared (IR) conformal fixed points. That is, if the conformal window is bounded below by a critical value $N_{fc}$, in some range of hyper-quark flavors $N_f$, we consider values of $N_f$ slightly below $N_{fc}$ to be outside the conformal window. We carry out a detailed analysis of this region in a generic bottom-up holographic setup, which includes both the dynamics of chiral symmetry breaking and the renormalization group (RG) flow.  To be specific, the RG flow can be modeled in holography by considering Einstein gravity coupled to a flowing dilaton field $\phi$. Then, as pointed out above, the quark degrees of freedom and the description of chiral symmetry breaking are added in terms of another scalar field $X$, dual to the $\bar qq$ operator. The interplay of the fields then gives rise to the picture expected from Dyson-Schwinger analysis, where the dimension of the operator $\bar q q $ flows from the free-field value, $\Delta_\mathrm{UV} = 3$ at the high-energy limit to $\Delta_\mathrm{IR} = 2$ near the IR fixed point.

In particular, this setup includes a complete model for the high-energy, ultraviolet (UV) physics, i.e., the flow from  an asymptotically free theory to the vicinity of the IR fixed point.
We do not discuss, however, a concrete model of the IR dynamics.
Instead, we consider a simple generic IR model imposed by boundary conditions to have a more general setup than the few explicit examples available in the literature.

Therefore, our setup is significantly more general than earlier studies in this class of models, which typically neglect the UV RG flow or chiral symmetry breaking or both, and restrict to certain models of the IR dynamics. This allows us to clarify which features of the setup are  model-independent. Note also that having a complete model of the UV flow enables us to analyze one and two-point functions consistently starting from the holographic dictionary.\\

To summarize:

\begin{itemize}

 \item We present a detailed analysis of the relevant one-point functions ($\theta_\mu^\mu$ etc.)
 \item We consider explicitly the effect of proper RG flow from the conformal IR fixed point to a UV free theory  (i.e. ``UV completion'')
 \item We do not discuss a concrete IR model (examples are given in literature) but do consider a simple generic IR model imposed by boundary conditions.
 \item This approximations mean that we explicitly consider a reasonable generic scenario for the UV region (but perhaps not some tuned exotic cases), i.e., basically everything that can happen in this region in a model-independent manner. For the IR, which is clearly model dependent, we impose a simple model which should essentially cover all possibilities. 
 \item We discuss explicitly how simpler holographic models, which only consider the physics near the IR fixed point, are embedded in these more generic models. 
\end{itemize}

It is believed that in walking theories, the intrinsic scale falls off exponentially with (the inverse square root of) $N_{fc}-N_f$ as the conformal window is approached from below. This is the so-called Miransky scaling \cite{Miransky:1996pd}. In this work, using our generic IR model, we find solutions which show the Miransky scaling law, but also such solutions, where scale separation between the IR and UV shows behaviors which are slightly modified with respect to the usual Miransky scaling, but still exponential. Interestingly, we find that light dilaton candidates can be present in such theories showing ``unusual'' scaling laws.

The rest of the article is organized as follows. In Section \ref{sec:setup} we introduce the general setup for the holographic model as well as a brief description of the interplay of the holographic RG flow and the chiral symmetry breaking. We also compute the PNGB dilaton mass and give general conditions for a light mode to be present. We also show that the masses of the pions, identified as the (flavored) phases of the field which is dual to the quark bilinear operator from our holographic model satisfy the Gell-Mann-Oakes-Renner relation. 
In Section \ref{sec:toymodel} we consider a simple toy model, which realizes in a concrete way most of the features of a more generic model and allows for analytic computations to be carried out explicitly. In this toy model we consider the flavor field $X$ in an AdS$_5$ background with constant dilaton potential, we also consider a varying mass of the flavor field to account for the scaling dimension of the bilinear operator, and varying boundary conditions as well. We treat the field $X$ in the probe limit and for the UV completion, we directly glue together solutions of $X$ at the IR and the UV. 
We analyze the possible outcomes for scale separation to see if it exhibits Miransky scaling, and the possibility of having a parametrically light scalar mode in the spectrum. In this section we also consider the effect of the hyper-quark mass on the scale hierarchy to find an Efimov spiral in the plane of the coefficients for the solution of the flavor field in the UV.
In Section \ref{sec:running} we consider a  more general case where we replace the cosmological constant in the AdS background of the previous toy model by a non-trivial dilaton potential to accommodate the holographic RG flow
of the dilaton that corresponds to the running of the coupling of the field theory. Thus we consider explicitly the effect of proper RG flow toward the conformal IR fixed point 
from a UV free theory (i.e. “UV completion”) until the dynamical generation of scale. We argue that the general model shows the same features as the simple analytic model of the preceding section. We also study the flavor contributions to the one-point and two-point functions of the Yang-Mills sector.

\section{General setup}\label{sec:setup}

Before defining specific models, let us discuss the general features of such bottom-up holographic frameworks for QCD-like theories we consider in this article.   
In general, we expect that the bulk five-dimensional action consists of gravity and matter sectors (closed- and open-string sectors)
\be \label{fullaction}
 S = S_\mathrm{grav}[g_{\m\n},\phi] + S_\mathrm{matter}[g_{\m\n},\phi,X] \ .
\ee
Here the gravity action is interpreted to be dual to the Yang-Mills theory, and the matter action is dual to the quark or flavor sector of the theory~\footnote{Here and onwards, we use a term quark for a hyperquark in strongly coupled gauge theories of near conformal dynamics. }. In particular, the field $X^{ij}$ is dual to the quark bilinear operator $\bar q^jq^i$ where $i,j$ are the flavor indices. The matter action can include other fields such as the gauge fields but these will not be important to describe the background or fluctuations in the scalar sector, where light modes are potentially expected.

The gravity action can be taken to be 
\be \label{eq:gravact}
 S_\mathrm{grav}= \frac{1}{2\kappa^2} \int d^5x \sqrt{-\det g}\left[R - g^{MN}\partial_M \phi \partial_N \phi + V(\phi) \right] 
\ee 
For the metric, we choose the conformal coordinates,
\be\label{eq:metric}
 ds^2 = e^{2A(r)}\left(dr^2-dt^2+d\mathbf{x}^2\right) \ ,
\ee
where $r$ represents the coordinate for the 5th dimension.

One may consider configurations with full backreaction of the matter, $S_\mathrm{grav} \sim S_\mathrm{matter}$, or in the probe limit, $S_\mathrm{grav} \gg S_\mathrm{matter}$. However, for the analysis of this article, the most important region is where the field $X$ is small. Notice that this is expected to hold quite in general close to the UV, at which the dimension of the $\bar q q$ operator should flow to the free value $\Delta_\mathrm{UV} = 3$. Then the field $X$ will be described at the leading order by a quadratic Lagrangian, 
\bea \label{eq:matter}
 S_\mathrm{matter} &= -\frac{1}{2\kappa^2N_c}\int d^5x \sqrt{-\det g}\ \mathrm{Tr}\left[g^{MN}\partial_M X^\dagger \partial_N X - m_X(r)^2 X^\dagger X \right] &\\
 &\simeq -\frac{\ell^3}{2\kappa^2N_c}\int d^5x\  r^{-5}\ \mathrm{Tr}\left[r^2\partial_M X^\dagger \partial_N X - \ell^2 m_X(r)^2 X^\dagger X \right]\,, \nonumber
\eea
where the trace is over the flavor indices of $X$ and we take for the metric to have approximately the AdS$_5$ form $e^{2A(r)} \simeq \ell^2/r^2$  on the second line, which is the expected asymptote near the UV boundary. In the following, we will consider flavour independent backgrounds so that $X$ will be proportional to the unit matrix in flavor space, unless stated otherwise.
Choosing the value of $m_X$ such that the scaling dimension of the quark bilinear at UV, $\Delta_\mathrm{UV}=3$, the UV asymptote for $X(r)$ takes schematically the following form
\be \label{eq:Xuvas}
 X(r \ll r_\uv) \sim m_q r + \sigma 
 r^3 \ ,
\ee
where $m_q$ is identified as the quark mass, $\sigma$ is proportional to the condensate $\langle \bar q q\rangle$
and $r_\uv$ is a certain scale as we will define in subsequent discussions.
Note that this formula implies that the field $X$ indeed becomes small near the boundary, which in turn also guarantees that it is enough to consider the flavor action in the probe limit. However, as we shall soon review, our analysis is not limited to the region where this asymptotic formula holds, but is valid for a much larger range of $r$ in the region of near-conformality.

We will make all of this explicit below by considering concrete models in following sections.

\subsection{Holographic description of the RG flow} \label{sec:RGflow}

We first remind (without complete derivations) what the standard picture of the RG flow for chiral symmetry breaking is and also what Miransky scaling near the conformal transition is. Here we consider the case of zero quark mass; in the rest of the article, we will also discuss flows with finite quark mass. As a function of the number of flavors; which we treat as a continuous parameter, the expected phases are the following: 
\begin{enumerate}
\item  $N_f>N_f^{\rm IR}$ (Coulomb phase). For a number of fermion flavors larger than a certain value, $N_f^{\rm IR}$, which is $3N_c$ for ${\rm SU}(N_c)$ this phase is described by a gauge theory with fermions in the fundamental representation, the asymptotic freedom is lost and the theory remains to be weakly coupled at low energy. We do not discuss the holographic implementation of the Coulomb phase in this article.

 \item $N_f^{\rm IR} > N_f > N_{fc}$ (conformal window). This phase is described by a chirally symmetric theory and with a flow ending in the IR fixed point, known as Banks-Zaks fixed point. For the chirally symmetric theory we have that $X=0$. Then the RG flow of the dilaton is simply between the two fixed points and there is a single scale, generated by the dimensional transmutation,  that characterizes the flow that we can identify as (the inverse of) $r_\mathrm{UV}$. For $r \ll r_\uv$ we have that $\phi(r) \approx \phi_\uv$, and for $r \gg r_\uv$ we have $\phi(r) \approx \phi_\ir$. At the IR fixed point, the IR  scaling dimension of the quark bilinear lies within the Breitenlohner-Freedman (BF) bound, $2<\Delta_\mathrm{IR}<3$. 
 \item $N_f = N_{fc}$ (critical point). As in the conformal window, in particular chiral symmetry stays intact, but now $\Delta_\mathrm{IR} =2$ at the IR fixed point so that the fluctuations of the flavor field are marginal. This means saturating the BF bound for the flavor field at the fixed point that defines the boundary of the conformal window or the conformal edge.
 
 \item $N_f < N_{fc}$ and $N_{fc}-N_f \ll 1$ (walking regime with Miransky scaling). This is the regime that we are mostly interested in as the theory is very close to the conformal window, having a very small $\beta$ function in the IR. The BF bound is violated, indicating an IR instability, which leads to chiral symmetry breaking, implemented in holography by non-zero $X$ field.
 There will be two scales, which are widely separated, and we denote the corresponding coordinate values by $r_\uv$ and $r_\ir (\gg r_\uv)$, and they will be related through the Miransky scaling law. There are three regimes for the solution
 \begin{itemize}
  \item The UV regime, $r \ll r_\uv$. The dilaton flows to the UV fixed point, $\phi(r) \approx \phi_\uv$, corresponding to the fact that the boundary theory is asymptotically free. The dimension of the chiral condensate runs to $\Delta_\mathrm{UV}=3$ in the UV, which controls the UV asymptotics of $X$.
  \item The walking regime,  $r_\uv \ll r \ll r_\ir$. This is the most interesting regime for us which encodes the model independent features of breaking of conformal symmetry, and which maps to the simplified analysis carried out above. The dilaton is already close to the IR fixed point value, $\phi(r) \approx \phi_\ir$. Therefore the dimension of the chiral condensate takes the fixed point value, $\Delta_\mathrm{IR} = 2 \pm i \nu$ (with $\nu \ll 1$) indicating the violation of the BF bound. This gives rise to a slowly oscillating solution for the $X$ field. The smallness of $X$ in this regime however guarantees that its solution is decoupled from the metric (as we assume above).
  \item The IR regime, $r \gtrsim r_\ir$. This is the model dependent region where the dilaton and $X$ are non-trivially coupled. Chiral symmetry breaking causes the flow of the dilaton to deviate from the fixed point value. The solutions depend on the chosen IR dynamics, such as the dilaton potential,  and/or IR boundary conditions.
 \end{itemize} 

\item $0< N_f < N_{fc}$ and $N_{fc}-N_f \sim {\cal O}(1)$, the QCD-like regime. In this regime the dynamics is similar to real QCD, i.e., the gauge theory is gapped, and at zero quark mass, there are massless pion states. When $N_{fc}-N_f$ grows to be order of one, the scale separation between the UV and IR disappears. In holography, the flow is therefore directly from the model-dependent IR regime to the weakly coupled UV fixed point.
 
\end{enumerate}

We stress that the precise behavior near the conformal transition at $N_f=N_{fc}$ is not known, and what we describe above is the standard expectation which arises e.g. from Dyson-Schwinger analysis and gauge/gravity models~\cite{Kaplan:2009kr,Jarvinen:2009fe,Jarvinen:2011qe,Alvares:2012kr}.

Let us also recall how the BF bound violation can give rise to the Miransky scaling. In the walking regime, $r_{\rm UV}<r<r_{\rm IR}$, we have the scaling dimension $\Delta_\mathrm{IR}=2\pm i \nu$ for the quark bilinear, which means that the solution for $X$ in the walking regime is given as $X \sim C_1 r^{2+i\nu} + C_2 r^{2-i\nu}$, or more conveniently, using real functions, as
\be \label{eq:Xwalkingsol}
 X(r) = X_0 \left( \frac{r}{r_\uv}\right)^2 \sin\left(\nu \ln\frac{r}{r_\uv}+\alpha\right) \ , 
\ee
where $X_0$ and $\alpha$ are constants.
One can argue that smoothness of the complete solution at $r \approx r_\uv$ and $r \approx r_\ir$ requires that these points are close to the nodes of the tachyon. The stability of the solution requires that there are no additional nodes between $r_\uv$ and $r_\ir$; this is the case because in this region where the field $X$ is small, its fluctuations would have essentially the same solution, i.e., a node in the wave function at zero mass, which indicates an instability. The scaling solution is therefore found if $r_\uv$ and $r_\ir$ are set by consecutive nodes. This gives
\be \label{eq:phasediff}
\nu\log(r_\ir/r_\uv) = \pi 
\ee
so that
\be \label{eq:Miranskyscaling}
 r_\ir = r_\uv \exp(\pi/\nu)
\ee
i.e., we obtained the expected Miransky scaling law as we approach the critical point $N_f=N_{fc}$, i.e., as  $\nu \to 0$ from above.
Notice that there is no reference to the beta function behavior in this derivation except that the theory is close to conformality. We will demonstrate these results in concrete models below. Note however, that our analysis below is not limited to models showing the ``standard'' Miransky scaling of~\eqref{eq:Miranskyscaling}, but we will also discuss more general scaling laws where the phase difference of~\eqref{eq:phasediff} deviates from $\pi$.
 

\subsection{Holographic computation of the dilaton mass} \label{sec:dilaton_mass}

One of our primary concerns is the holographic computation of the mass of the PNGB (Pseudo-Nambu-Goldstone Boson) dilaton arising from the approximate conformal symmetry in the walking regime. In our previous work~\cite{CruzRojas:2023jhw}, it has been noticed that a parametrically light PNGB dilaton, compared to the IR conformal symmetry breaking scale, can be analytically obtained assuming small backreaction from the flavor field $X$. Here we summarize the argument for this and improve the previous discussion. Note that the potential light mode is found in the flavor sector, i.e., it is a meson mode rather than a glueball. In the holographic context this is natural as it is the flavor field $X$ that controls the scale separation and walking, as we discussed above. There is no analogous mechanism in the gluon sector.\footnote{In Sec.~\ref{sec:flucts}, we will show explicitly that the mixing between the meson mode and the glueball mode is suppressed in the walking regime (i.e. $r_\uv \ll r \ll r_\ir $). In principle, the mixing can be non-negligible in the deep IR regime ($r \gtrsim r_\ir$). However, in our analysis, we assume that the IR physics can be parameterized by a boundary condition near the edge of the walking regime, after cutting off the IR geometry ($r \lesssim r_\ir$), so that the mixing is still negligible, and the PNGB dilaton is identified as a meson mode.} This is in agreement with several earlier holographic studies, where mesons are typically the lightest states even if they were not parametrically light (see, e.g.,~\cite{Kutasov:2011fr,Arean:2013tja,CruzRojas:2023jhw}). Also lattice analyses of near-conformal theories have found signs of such light states~\cite{Brower:2015owo,LatKMI:2016xxi,LatticeStrongDynamics:2018hun}.

In the probe limit, we may directly study the fluctuation of the $X$ field, $\chi$. The corresponding gauge invariant field variable $\xi(x^M)$ and its equations are given in appendix~\ref{app:fluctuations} and for the homogeneous background of~\eqref{eq:Xwalkingsol} it has a solution in terms of Bessel functions
\be \label{eq:Besselsol}
 \xi(r) =  \frac{C_1 \Re [J_{i\nu}(\omega r)]+C_2 \Re[ Y_{i\nu}(\omega r)]}{2 \sin(\nu \ln \frac{r}{r_\uv}+ \alpha) + \nu \cos (\nu \ln \frac{r}{r_\uv} + \alpha) } \ ,
\ee
where $\omega$ is the mass of the fluctuation mode and $\nu \ll 1$ as required for walking solutions.
Since we are interested in light modes, let us consider small $\omega$. For small $\omega$, the above solution is approximated as
\dis{ \label{xiappsol}
 \xi(r) \simeq & \left[ C_1 \left(\cos z_\nu -\frac{\omega^2 r^2}{4(1+\nu^2)}\left(\cos z_\nu + \nu \sin z_\nu\right) \right) \right.\\  
 & \left. + C_2  \left(\frac{2}{\pi \nu} \sin z_\nu  +\frac{\omega^2 r^2}{2\pi \nu (1+\nu^2)} \left(\nu \cos z_\nu - \sin z_\nu\right)\right) \right] \\
 &\times \frac{1}{2 \sin(\nu \ln \frac{r}{r_\uv}+ \alpha) + \nu \cos (\nu \ln \frac{r}{r_\uv} + \alpha) } \,,
}
where 
\dis{
z_\nu \equiv \nu \ln (\omega r/2) - \gamma_\nu\, , \label{znu}
} 
and $\gamma_\nu$ is the complex phase of the gamma function 
\dis{\label{eq:gammanudef}
\Gamma(1+ i \nu) = e^{i \gamma_\nu} \sqrt{\pi \nu/\sinh(\pi \nu)}} 
so that $\gamma_0 = 0$.

In order to describe the physics near $r=r_\ir$ in a model-independent way, we introduce a generic, but linear, IR boundary condition with two constant parameters $\mathcal{A}$ and $\mathcal{B}$~\footnote{Here we are grossly simplifying the IR dynamics with two parameters, $r_{\rm IR}$ and the ratio, ${\cal A}/{\cal B}$. }.
\be
 \mathcal{A}\, \xi(r_\ir) + \mathcal{B}\, r_\ir \xi'(r_\ir) = 0, \label{IRbc}
\ee
where the prime in $\xi'$ denotes differentiation with respect to the coordinate $r$.
For the approximate solution (\ref{xiappsol}), this IR boundary condition amounts to
\dis{ \label{ircond}
&\left(\frac{\mathcal{A}}{\mathcal{B}} +\nu \frac{2  + \nu \tan(\beta-\alpha)}{2 \tan(\beta-\alpha) - \nu} \right) 
\left[ \frac{C_1}{C_2} \left(1-\frac{\omega^2 r_\ir^2}{4}\right)+\frac{1}{\pi \nu}\left( 2t_{z_\nu^{\rm IR}}+\frac{\omega^2 r_\ir^2}{2}\left( \nu - t_{z_\nu^{\rm IR}}\right)\right) \right]  \\
&\quad\quad\quad\quad\simeq \Bigg[ \frac{C_1}{C_2} \left(\nu t_{z_\nu^{\rm IR}}+\frac{\omega^2 r_\ir^2}{2}\right)-\frac{2}{\pi}\left( 1+\frac{\omega^2 r_\ir^2}{2}\left[ \frac12 - \frac{1}{\nu} t_{z_\nu^{\rm IR}}\right]\right) \Bigg] 
}
where $t_{z_\nu^{\rm IR}}= \tan({z_\nu^{\rm IR}})$, $z_\nu^{\rm IR} \equiv z_\nu|_{r=r_\ir}$, and $\beta$ is defined by
\dis{ \label{eq:betadef}
\frac{r_\ir}{r_\uv} \equiv e^{(\pi - \beta)/\nu}
}
with $0<\alpha<\beta <\pi$ for stability of the background solution for the flavor field $X$.

As for UV boundary condition, on the other hand, we may require for the UV normalizability of the solution
\dis{
 \xi(r_\uv) \equiv \xi_\uv \ll 1. \label{xiuvbc}
 }
This gives us 
\be \label{uvcond}
\frac{C_1}{C_2} \simeq 
-\frac{2}{\pi \nu} (\tan z_\nu^{\rm UV} + c_\uv)
\,,
\ee
where 
\bea
\label{znuIRUV}
z_\nu^{\rm UV} &\equiv& z_\nu|_{r=r_\uv} = z_\nu^{\rm IR}+\beta -\pi\,,\\
c_{\rm UV} &\equiv&-\frac{\pi \nu}{2} (2 s_\alpha + \nu c_\alpha) \frac{\xi_{\rm UV}}{C_2 \textrm{Re}[J_{i\nu}(\omega r_{\rm UV})] } \label{cuv}
\eea
with $\sin\alpha\equiv s_\alpha$ and  $\cos\alpha\equiv c_\alpha$.

Now combining the IR condition (\ref{ircond}) and the UV condition (\ref{uvcond}),
\dis{
\omega^2 r_\ir^2 \simeq 4 \frac{\left(\frac{\cal A}{\cal B} + \nu  \frac{1+\frac\nu2 t_{\beta-\alpha}}{t_{\beta-\alpha}-\frac\nu2}\right) (t_{z_\nu^\ir+\beta}-t_{z_\nu^\ir}+c_\uv) -\nu\left(1+ t_{z_\nu^\ir}(t_{z_\nu^\ir +\beta}+c_\uv) \right)}{\left(2+\frac{\cal A}{\cal B} + \nu \frac{1+\frac\nu2 t_{\beta-\alpha}}{t_{\beta-\alpha}-\frac\nu2} \right)(t_{z_\nu^\ir+\beta}-t_{z_\nu^\ir}+c_\uv) } \label{lmass}
}
where $t_\beta$ stands for $\tan \beta$. 
Here one can immediately see that a necessary condition for a light mode 
$\omega \ll r_\ir^{-1}$ is ${\cal A}=0$ (Neumann IR boundary condition) or ${\cal A}/{\cal B} \ll 1$. Thus taking ${\cal A}=0$ and further assuming $c_\uv\simeq 0$ from the UV normalizability condition (\ref{xiuvbc}), Eq. (\ref{lmass}) becomes
\dis{
\omega^2 r_\ir^2 = \frac{2 \nu \sin \alpha}{\sin(\beta-\alpha) \sin \beta} + \frac{\nu^2}{\sin^2 \beta} + {\cal O}(\nu^3)
}
Therefore, a parametrically light dilaton is obtained if $\beta \gg \nu$, indicating a (small) deviation from the Miransky scaling (see~\eqref{eq:betadef}).


\subsection{Pions and the Gell-Mann-Oakes-Renner relation}

Let us then discuss the masses of the pions and in particular show that they satisfy the Gell-Mann-Oakes-Renner relation at small quark mass. It is well known that holographic models with spontaneously broken chiral symmetry produce this relation with weak assumptions~\cite{Evans:2004ia, Sakai:2004cn,Bergman:2007pm}. Here we sketch how the analysis works in our case.

The pions are identified as the (flavored) phases of the $X$ field. However, in order to analyze them correctly, one needs to gauge the action~\eqref{eq:matter}, i.e., promote the $SU(N_f)_L\times SU(N_f)_R$ chiral symmetry of QCD to a gauge symmetry on the gravity side following the standard gauge/gravity duality dictionary. That is, we write
\begin{align}
 \label{eq:pionL}
 S_\mathrm{matter} &= -\frac{1}{2\kappa^2N_c}\int d^5x \sqrt{-\det g}\ \mathrm{Tr}\bigg[g^{MN}D_M X^\dagger D_N X - m_X(r)^2 X^\dagger X  \nonumber\\
 & \quad + \frac{w(r)}{4}F^{(L)}_{MN} F^{(L)MN}+ \frac{w(r)}{4}F^{(R)}_{MN} F^{(R)MN} \bigg]
\end{align}
where 
\be
 D_M X = \partial_M X - i X A_{M}^{(R)}  - i  A_{M}^{(L)} X\ ,
\ee
and $F^{(L/R)}$ are the field strength tensors of the gauge fields $A_{L/R}$. Note that this is the unique gauge-invariant choice of action up to second-order in the fields.

The dependence on the holographic coordinate in the functions $m_X$ and $w$ in~\eqref{eq:pionL} is expected to arise through additional scalar fields. However, it turns out that for the analysis of the pion modes it is enough to consider the fluctuations near the boundary, $r \to 0$. In this region, as above, we expect that $m_X^2 \approx 3/\ell^2$, fixed by the dimension of the dual operator $\bar q q$, while $w$ tends to a constant whose value is not relevant for the pion mode. Moreover, the geometry takes the AdS$_5$ form. 

The phases of $X$ couple to the longitudinal modes of the axial gauge field. Therefore we turn on the fluctuations for the phases
\be
 X \mapsto X e^{2it^a\hat \pi^a}
\ee
where $\hat \pi^a(r,x) = \pi^a(r)e^{-ix_\nu q^\nu}$ with $t^a\, (a=1, \dots, N_f^2-1)$ being the generators of $SU(N_f)$ the fundamental representation and $X$ is proportional to unit matrix in flavor space with flavor-independent quark masses for simplicity. We assume the normalization $\mathrm{Tr}(t^at^b) = \delta^{ab}/2$. Moreover, we write 
\be
 A^{(L)}_\mu(r,x) = - A^{(R)}_\mu(r,x) = i q_\mu a^b(r)e^{-ix_\mu q^\mu} t^b
\ee
for the gauge fields. We use the standard radial gauge, where the $r$-components of the gauge fields vanish. Then the fluctuation equations are
\begin{align}
\label{eq:longflucts}
\begin{aligned}
 r\frac{d}{dr}\left[\frac{1}{r}a'(r)\right]-\frac{4X(r)^2\ell^2}{w\, r^2}\left[a(r)-\pi(r)\right]  &= 0& \\
 \frac{4X(r)^2\ell^2}{r^2} \pi'(r) - m^2 w\, a'(r) &= 0&
\end{aligned}
\end{align}
where $m^2 = q^2$ is the mass of the fluctuations. Here we have hidden the flavor indices as the equations are diagonal in flavor space.

Pion fields couple to the longitudinal terms in the axial currents $J_{\mu 5}^a = \bar q \gamma_\mu\gamma_5 t^a q$. By its definition, the pion decay constant is found as the residue of the longitudinal correlator:
\be 
 -iq_\mu q_\nu \int d^4x\, e^{-iq_\alpha x^\alpha}\langle 0|\,T\left\{J^{\mu a}_5(x)J^{\nu b}_5(0)\right\}| 0\rangle \approx \delta^{ab} \frac{m_\pi^4f_\pi^2}{q^2-m_\pi^2} \ ,\qquad (q^2 \to m_\pi^2) \ .
\ee
In holography, the longitudinal correlator is computed by considering the fluctuations in~\eqref{eq:longflucts} with a boundary value for the field $a$. A standard computation (see, e.g.,~\cite{Erlich:2005qh}) gives the residue in terms of the corresponding normalizable fluctuation wave function at the boundary. In this case the normalizable mode is the pion mode $a_0$, $\pi_0$, and the decay constant can be obtained from the near-boundary behavior of the gauge-field component:
\be \label{eq:fpiholo}
 m_\pi^2 f_\pi^2 = \frac{2}{\kappa^2 N_c}\lim_{r\to 0}\left( \frac{\sqrt{w}\ell a_0'(r)}{r}\right)^2 \ .
\ee
Here normalization was chosen such that the $a_0$ component satisfies the condition~\cite{Jarvinen:2015ofa} 
\be \label{eq:longnorm}
 1 \approx \int_0^{r_\mathrm{cut}} dr \frac{w\, r}{\ell X(r)^2} \left(a_0'(r)\right)^2 = \int_0^{r_\mathrm{cut}} dr \frac{r^3}{X(r)^2\ell^3} \left( \frac{\sqrt{w}\ell a_0'(r)}{r}\right)^2 \ ,
\ee
where the upper bound of the interval $r_\mathrm{cut}$ should be within the range where our near-boundary approximations work, but its precise value turns out to be irrelevant.
Namely, analyzing the fluctuation equations~\eqref{eq:longflucts}, for normalizable modes the combination appearing both in~\eqref{eq:fpiholo} and in~\eqref{eq:longnorm} indeed goes to a constant near the boundary,
\be
 \frac{\sqrt{w}\ell a_0'(r)}{r} = a_c + \morder{r^2} \ .
\ee
As the $X$ field obeys the asymptotics~\eqref{eq:Xuvas} near the boundary, it follows that the integral in~\eqref{eq:longnorm} is dominated at $r \sim \sqrt{m_q\sigma}$, which tends to zero as $m_q \to 0$. We therefore find
\be
 1 \approx a_c^2 \int_0^{r_\mathrm{cut}} dr \frac{r^3}{X(r)^2\ell^3} \approx a_c^2 \int_0^{\infty} dr \frac{r^3}{\left(m_q r + \sigma r^3 \right)^2\ell^3} = \frac{a_c^2}{2m_q \sigma \ell^3} \ .
\ee
Inserting the obtained value of $a_c$ in~\eqref{eq:fpiholo} gives the Gell-Mann-Oakes-Renner relation\footnote{Here note that we assume the flavor-independent background $X$ with flavor-independent quark mass $m_q$.
}
\be \label{eq:gmor1}
  m_\pi^2 f_\pi^2 \approx \frac{4}{\kappa^2 N_c} \ell^3 m_q \sigma \ .
\ee
The relation of $\sigma$ to the chiral condensate may be computed by following the holographic dictionary (see Appendix~\ref{app:holo_renorm}):
\be
 -\frac{2 N_f  \ell^3}{\kappa^2 N_c}\sigma = \sum_{i=1}^{N_f}\langle\bar q_i q_i\rangle  = N_f \langle\bar q_i q_i\rangle\ ,
\ee
where in the last expression there is no sum over the flavor indices.
Inserting this in~\eqref{eq:gmor1} gives
\be
 m_\pi^2 f_\pi^2 \approx - 2m_q\langle\bar q_i q_i\rangle
\ee
where the quark condensate $\langle \bar{q}_i q_i \rangle$ is actually flavor-universal.

Note that this result only refers to the behavior of the background near the boundary. The behavior in the IR region, in particular whether there is walking or not, is irrelevant.  


\section{Toy model}\label{sec:toymodel}

It is useful to first consider a simple model which realises most of the features of a more generic model, but allows for completely analytic solutions of the background so that many observables can be computed explicitly. We will refer to this model as the ``toy model'' below.

We consider the flavor field $X$ in an (exact) AdS$_5$ background i.e., essentially the choice of the dilaton potential $V(\phi)=\Lambda_5$ in~\eqref{eq:gravact}, with varying mass of the flavor field and varying boundary conditions.

The explicit definition for the matter action is given in (\ref{eq:matter}), and additionally we set the following:
\begin{itemize}
 \item We treat the field $X$ in the probe limit, assuming only quadratic action also in the IR, and set generic IR boundary conditions:
 \be \label{eq:IRbcgen}
A X(r_\ir) + B r_\ir X'(r_\ir) = 0 \ ,
\ee
where $r_{\rm IR}$ is the IR scale around which the walking behavior ceases, and  $A$ and $B$ are constants. The idea is that the IR dynamics can be complicated, such as in known examples~\cite{Jarvinen:2011qe,Pomarol:2019aae} and is likely to set boundary conditions that are essentially random and independent of the solutions near the boundary. We argue in Appendix~\ref{app:bcs} that these simple boundary conditions can indeed model accurately more generic boundary conditions when the UV and the IR scales of the theory are well separated.

 \item For the UV completion, we glue together directly solutions of $X$ with $\Delta_\ir =2 \pm i\nu$ and $\Delta_\uv=3$, requiring continuity of the function and its first derivative.
It means
 \be \label{mX}
- \ell^2 m_X(r)^2 = \left\{
 \begin{array}{clc}
     \Delta_\mathrm{UV}(4-\Delta_\mathrm{UV}) & = 3 & \qquad (r<r_\mathrm{UV})    \\
     \Delta_\mathrm{IR}(4-\Delta_\mathrm{IR}) & = 4 + \nu^2 &  \qquad (r_\mathrm{UV} < r < r_{\rm IR})
 \end{array}
 \right.
\ee
where $r_\mathrm{UV}$ is a cutoff value, to be associated with the UV scale of the theory. This toy model is presumably realised through a dilaton potential $V(\phi)$ with a rapid step, and it works as a model for the expected behavior (discussed above) where the theory flows between two fixed points, and is analytically tractable. In section \ref{sec:running}, we will argue that the characteristic features of generic UV completions are similar to this toy model, even though the details differ.
\end{itemize}

We are mostly interested in the case where $\nu$ is taken to be small, even though the model is also well defined at finite and positive $\nu$. Moreover, we shall take the matrix $X$ to be proportional to the unit matrix $X^{ij}\propto \delta ^{ij}$, as we are mainly interested in the flavor singlet sector. We will not write the indices explicitly in the following.

\subsection{Scale hierarchy}

Let us first discuss implications of the toy model on the hierarchy between the UV scale $r_{\rm UV}$ and the IR scale $r_{\rm IR}$.
From (\ref{mX}) the solution for $X(r)$ is given by
\begin{align}
 X(r) &= m_q r + \sigma r^3 \ ,&  \  &(r<r_\uv)& \label{eq:XUVtoy1} \\
  X(r) &= X_0 \left(\frac{r}{r_\uv}\right)^2 \sin\left(\nu \log \frac{r}{r_\uv}+\alpha\right) \ ,&  \  &(r_\uv<r < r_\ir)& \label{eq:XIRtoy}
\end{align}
where we included the factor of $1/\nu$ in the normalization of the IR solution for convenience. 

As for the UV boundary condition at $r=r_{\rm UV}$, we can directly match the IR solution~\eqref{eq:XIRtoy} with the UV solution~\eqref{eq:XUVtoy1}. That is, we glue these solutions together at $r=r_\mathrm{UV}$, requiring the continuity of the function and its first derivative. It determines the phase, $\alpha$, and the field value, $X_0$, around the UV scale in terms of $m_q$ and $\sigma$ as
\bea 
 \tan\alpha &=& \nu \frac{\sigma r_\uv^3 + m_q r_\uv}{\sigma r_\uv^3 - m_q r_\uv}, \label{alpha_toy}\\
 X_0 \sin \alpha &=& \sigma r_\uv^3 + m_q r_\uv, \label{X0_toy}
\eea
where $-\pi/2<\alpha<\pi/2$ without loss of generality.

On the other hand, the IR boundary condition (\ref{eq:IRbcgen}) determines the IR scale $r_\ir$ in terms of the ratio $A/B$ as
\bea
\log \frac{r_\ir}{r_\uv}&=-\frac{\alpha}{\nu} -\frac{1}{\nu} \tan^{-1} \left( \frac{\nu}{2+A/B}\right), \label{scale} \\
&\simeq -\left(\frac{\alpha}{\nu}+ \frac{1}{2+A/B}\right) + \frac{n\pi}{\nu}, \label{scale_a}
\eea
where $n$ is an integer and we assume $2+A/B \gg \nu$ for the second line. Since $r_\ir > r_\uv$, if the quantity in the parenthesis is negative, it allows us to have a solution with $n=0$, yielding a non-Miransky scaling. If that quantity is positive, $n$ has to be equal to or greater than 1. 
For $n=1$, we may have approximately Miransky scaling, i.e. $\log (r_\ir/r_\uv) \approx \pi/\nu$, assuming that $\alpha \sim \nu$ as~\eqref{alpha_toy} suggests. On the other hand, $n>1$ implies that the solution $X(r)$ has a node where the background field $X$ is vanishing. Such a vanishing background is likely to yield an instability of the solution upon perturbation so that the system eventually goes to the $n=1$ solution. We will show this more precisely later by computing free energy of each configuration. Note that the presence of a tower of unstable condensed solutions is not specific to our setup but a rather general consequence of BF bound violation (see, e.g.,~\cite{Kutasov:2011fr,Iqbal:2011in}).

The above argument shows that, generally, both Miransky scaling and non-Miransky scaling are possible depending on the IR physics parameterized by the parameter $A/B$. 
Let us look at more details separately for the cases with $m_q=0$ and $m_q\neq0$ in the following subsections. 

\subsubsection{Massless quarks}

Let us first consider the case that the quarks are massless. Eq. (\ref{alpha_toy}) with $m_q=0$ gives us 
\dis{
\alpha = \tan^{-1} \nu \simeq \nu
}
so that $\alpha$ has no dependence on the UV condensation $\sigma$.
Thus the scale hierarchy (\ref{scale}) is solely determined by the model parameters $\nu$ and $A/B$ as
\be
 \log\frac{r_\ir}{r_\uv} = \frac{\pi n}{\nu} - \frac{A/B+3}{A/B+2} +\morder{\nu}.
\ee
Consistency requires that $r_\ir\ge r_\uv$. Therefore the solution with $n=0$ is possible only when $(A/B+3)/(A/B+2)\le 0$, i.e., $-3 \le A/B<-2$. In this narrow range of IR boundary conditions,  $n=0$ gives the ground state solution, $r_\ir \sim r_\uv$, unless $A/B$ is very close to $-2$, and there is no Miransky scaling. For the $n \geq 1$ solution with $A/B$ in this range, the background field $X$ has a node indicating instability.
For other values of $A/B$, the solution with $n=0$ is not possible. Then $n=1$ solution is the only solution without a node, and it obeys Miransky scaling in the limit $\nu \to 0$. In particular, Miransky scaling is found both with Dirichlet ($B=0$) and Neumann ($A=0$) boundary conditions for $X$. 

However, we can also take $\nu$ to be finite but small, instead of taking $\nu \to 0$ directly. Then close to the critical values of $A/B$, where the scaling changes, a special solution can be found.
In particular, the case where $A+2B=0$ exactly is special. Then Eq. (\ref{scale}) tells us 
\be \label{eq:onehalfscaling}
 \log\frac{r_\ir}{r_\uv} = \frac{\pi }{2\nu} -\frac{\alpha}{\nu} \simeq \frac{\pi}{2\nu}
\ee
in order to not have a node in the solution of $X(r)$. 
The scaling is similar to the Miransky scaling but with a numerical factor smaller by a factor of $2$.
This special case is also interesting since
\be
 X(r) \approx X_0 \left(\frac{r}{r_\uv}\right)^2\ , \qquad \nu \to 0
\ee
near $r= r_\ir$ so that there is no log term in the background. 

Other numerical factors than $1/2$ in the scaling law~\eqref{eq:onehalfscaling} may be obtained by taking $A+2B= \nu \kappa B$ and then taking the limit $\nu \to 0$: in this case Eq. (\ref{scale}) becomes
\be
  \log\frac{r_\ir}{r_\uv} = \frac{\pi n}{\nu} - \frac{1}{\nu}\tan^{-1} \frac{1}{\kappa} -\frac{\alpha}{\nu}
\ee
The ground state has $n=1$ ($n=0$) for positive (negative) $\kappa$. The solution can be written as
\be
  \log\frac{r_\ir}{r_\uv} \simeq \frac{\pi-\cot^{-1}\kappa}{\nu}
\ee
where the branch of the arccot function was chosen such that the values of the function are between $0$ and $\pi$.

\subsubsection{Massive quarks}

In order to study the quark mass effect on the scale hierarchy, we may compare the quark mass with the IR dynamical mass $X(r_\ir)$. 
Let us define
\bea
\hat{m}_q &\equiv& \frac{m_q r_\uv}{X_0^\ir} =\left(\frac{r_\uv}{r_\ir} \right)^2 \frac{1}{\nu}(\sin \alpha - \nu \cos \alpha)\,, \label{mqh} \\
\hat{\sigma} &\equiv& \frac{\sigma r_\uv^3}{X_0^\ir} = \left(\frac{r_\uv}{r_\ir} \right)^2 \frac{1}{\nu}(\sin \alpha + \nu \cos \alpha) \,, \label{sigmah}
\eea
where 
\dis{
X_0^{\rm IR} \equiv \nu X_0 \left(\frac{r_\ir}{r_\uv}\right)^2 = \frac{\nu}{\sin\left(\nu\ln\frac{r_\ir}{r_\uv} + \alpha \right)} X(r_\ir).
}
Here Eq. (\ref{mqh}) and Eq. (\ref{sigmah}) are obtained from Eq. (\ref{alpha_toy}) and Eq. (\ref{X0_toy}). By Eq. (\ref{scale}) let us also write 
\dis{
\alpha= -\nu \ln \frac{r_\ir}{r_\uv}-\tan^{-1} \frac{\nu}{2+A/B}\,.\label{alphah}
}

\begin{figure}
\centering
 \includegraphics[width=0.7\textwidth]{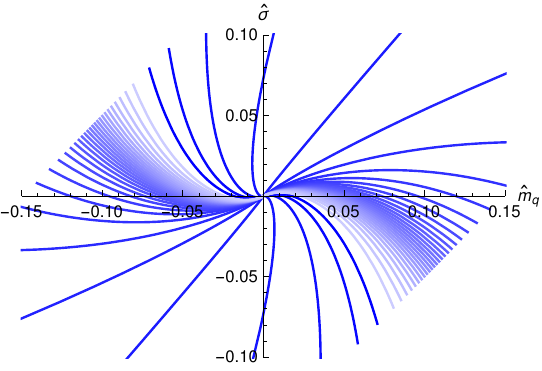}
 \caption{The Efimov spiral with varying boundary conditions. We chose $\nu = 0.15$ and $A/B = \tan \phi$ with $\phi$ varying from $-1.52$ to $1.48$ in steps of $0.1$. The lighter (darker) shades indicated lower (higher) values of $\phi$. 
 }
 \label{fig:spirallinear}
\end{figure}

\begin{figure}
\centering
 \includegraphics[width=0.9\textwidth]{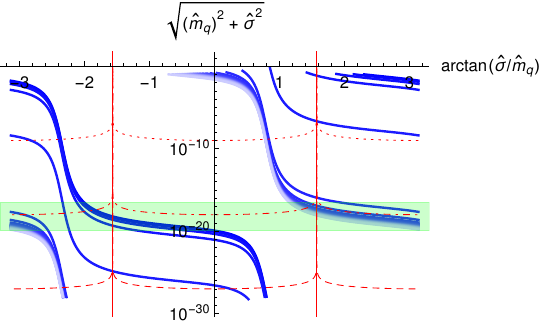}
 \caption{The Efimov spiral with varying boundary conditions in polar coordinates. Notice the logarithmic scaling of the horizontal axis. Notation of the spiral (blue curves) as in Fig.~\protect\ref{fig:spirallinear}, expect that for clarity we only show one of the spirals related by the change of sign $(\hat m_q,\hat \sigma) \mapsto (-\hat m_q,-\hat \sigma)$. The red solid, dashed, dotdashed, and dotted curves are the curves of $|\hat m_q| = 0$, $10^{-27}$, $10^{-18}$, and $10^{-9}$, respectively. The shaded green band shows the region of Miransky scaling (allowing a 10\% deviation from the precise value).}
 \label{fig:spirallog}
\end{figure}

Using Eq. (\ref{mqh}), Eq. (\ref{sigmah}), and Eq. (\ref{alphah}), we can plot all possible values of the ratios $\hat{m}_q$ and $\hat{\sigma}$ as the scale hierarchy $r_\ir/r_\uv$ varies. 
This will lead to a spiral in the ($\hat{m}_q,\hat{\sigma}$) plane which has been coined the Efimov spiral~\cite{Efimov:1970zz,Iqbal:2011aj}. The spirals are shown for various choices of $A/B$ with $\nu =0.15$ in Figs.~\ref{fig:spirallinear} and~\ref{fig:spirallog}. 
Note that such spirals are common in holographic models (see, e.g.,~\cite{Iqbal:2011in,Iqbal:2011aj}). In a typical setting (similarly as in this work), the spirals are driven by the violation of the BF bound at a fixed point. In models similar to the current setup, spirals have been considered in~\cite{Jarvinen:2015ofa,Arean:2016hcs,BitaghsirFadafan:2018efw}.

The spiral structure of Eqs. (\ref{mqh}) and (\ref{sigmah}) can also be made explicit.
Note that above we required that $0<\alpha<\pi$. If we allow this parameter to take any real values, the branch ambiguity of $\tan^{-1}$ in~\eqref{alphah} may be absorbed in $\alpha$. That is, we define the parameter $\bar \alpha = \pi n -\alpha$ where the integer $n$ arises for the branches of $\tan^{-1}$ as above. In terms of this parameter, 
\bea
\hat{m}_q = \exp\left(\frac{2}{\nu}\tan^{-1}\frac{\nu}{2+A/B}\right)\exp\left(-\frac{2\bar\alpha}{\nu}\right) \frac{1}{\nu}(\sin \bar \alpha + \nu \cos \bar \alpha)\,, \label{mqhsp} \\
\hat{\sigma}  = \exp\left(\frac{2}{\nu}\tan^{-1}\frac{\nu}{2+A/B}\right)\exp\left(-\frac{2\bar\alpha}{\nu}\right) \frac{1}{\nu}(\sin \bar \alpha - \nu \cos \bar \alpha) \,, \label{sigmahsp}
\eea
with the understanding that a specific branch of $\tan^{-1}$ is picked, and overall signs are absorbed into the constant $X_0^\ir$ in~\eqref{mqh} and~\eqref{sigmah}. These equations indeed describe a spiral as the parameter $\bar\alpha$ varies. The physically meaningful solutions with $r_\ir>r_\uv$ are found for positive values of $\bar\alpha$.

Let us discuss the structure of the figures, starting from Fig.~\ref{fig:spirallinear}. We chose to express the ratio $A/B$ in terms of an angle, $A/B =\tan \phi$, and to vary the angle linearly. This choice does not set any of the boundary conditions to a special role. The colors vary from light blue (at $\phi=-1.52$, chosen to lie close to $\phi=-\pi/2$, i.e., Dirichlet boundary conditions) to dark blue (at $\phi=1.48$). Notice that for each value, there are actually two spirals, related by reflection with respect to the origin. Let us also note that
\dis{
\sqrt{\hat{m}_q^2 + \hat{\sigma}^2} =\frac{\sqrt{2}}{\nu} \left(\frac{r_\uv}{r_\ir} \right)^2 \sqrt{\sin^2\alpha + \nu^2 \cos^2 \alpha} \sim \left(\frac{r_\uv}{r_\ir} \right)^2\,,
}
which means the radius of the spiral measures the scale hierarchy squared. 
In the linear plot of Fig.~\ref{fig:spirallinear} the spiral structure is actually not well visible, because the radius $\sim r_\uv^2/r_\ir^2$ goes to zero exponentially fast with respect to the variation of the phase. In particular, only one regular solution at $\hat m=0$ (and its trivial reflected counterpart) can be identified. The other solutions at zero mass are so close to the origin that they cannot be distinguished from it. 

This is why we switch to polar coordinates, and plot the angle $\arctan(\hat\sigma/\hat m_q)$ as well as the radius $\sqrt{\hat m_q^2+\hat \sigma^2}$ in logarithmic scale in Fig.~\ref{fig:spirallog}. Now each spiral starts at the top of the plot, and evolves downwards in step while the angle increases. Notice that a curve that exits on the right enters  from the left at the same value of the radius. The structure continues to indefinitely small values of the radial coordinate, so that each curve indeed circles the origin infinitely many times. For clarity, unlike in Fig.~\ref{fig:spirallinear}, we only show one spiral for each value of $\tan \phi$.

The backgrounds with zero mass are found at the intersection points with the solid red lines. For each curve, i.e., each IR boundary condition, there are infinitely many such solutions, but only the first and in some cases second of the solutions is visible in the plotted area. We note that almost all such solutions show Miransky scaling with $\log\hat\sigma \approx -2\pi/\nu \approx -41.9$ (the light green band in the plot). However, one of the solutions shows essentially no scaling, and another shows non-standard scaling. This solution with non-standard scaling is the finite-$\nu$ counterpart of the critical boundary conditions with $A/B=-2$ -- actually for this curve we have $A/B \approx -2.066$.

Let us then comment on the behavior of small but finite quark mass. Examples of the curves with constant quark mass are given as the dashed, dotted, and dot-dashed curves in Fig.~\ref{fig:spirallog}. Notice that because signs of fermion masses are not observables, we include solutions with both positive and negative $\hat m_q$ in the curves. 
When the quark mass is tiny (e.g., $\hat m_q \ll \exp(-2\pi/\nu)$ for Miransky scaling solutions), it can be treated as small perturbation and the ratio of the UV and IR scales is essentially unaffected. An example of such a case is the red dashed curve in Fig.~\ref{fig:spirallog}: with this mass value the regular solutions are found at the intersection points of the red dotted curve with the blue curves. When $\hat m_q$ grows and starts to be of the same order with $\hat \sigma$  (e.g., $\hat m_q \sim \exp(-2\pi/\nu)$ for Miransky scaling solutions), it can be no longer treated as a small perturbation and the ratio of the UV and IR scales is affected non-trivially (but Miransky scaling still holds) for those IR boundary conditions which show Miransky scaling at zero mass. An example of such a case is the red dot-dashed curve in Fig.~\ref{fig:spirallog}. For even larger masses i.e., $\exp(-2\pi/\nu) \ll \hat m_q \ll 1$, the ratio between the UV and IR scales is determined by the quark mass, except for the few special cases which have non-standard scaling behavior. And example of such is shown as the red dotted curve. For the backgrounds with standard Miransky scaling, these results agree with the analysis of~\cite{Jarvinen:2015ofa}.

From Fig.~\ref{fig:spirallog} we also see that the solutions with non-Miransky scaling, in the parts of the curves that drop steeply, typically satisfy $\arctan(\hat\sigma/\hat m_q) \approx \pi/4$ (or $-3\pi/4$) meaning that $\hat m_q \approx \hat \sigma$. From Eq.~(\ref{alpha_toy}) we see that the UV phase $\alpha$ is large in this region, i.e., it deviates from the expected generic behavior $\alpha \sim \nu$. While these steep steps in the curves are expected to be universal features that are present also in more complicated models than the simple toy model, the location of the drops is model dependent. It is also possible, that the drop happens near $\arctan(\hat\sigma/\hat m_q) \approx \pi/2$ in some models with UV structure that differs from the toy model, which would mean that the non-standard scaling solutions are found even at small quark masses for typical IR boundary conditions.

From Fig.~\eqref{fig:spirallog} it is clear that for small values of the quark mass and fixed IR boundary conditions there are several regular solutions.
In order to determine which of the solutions dominates, one should compute their free energies, which is obtained as the value of the on-shell action. The free energy is divergent and needs to be renormalized: for the toy model, it is enough to add the counterterm given in~\eqref{eq:Xct} in Appendix~\ref{app:holo_renorm}. Direct computation of the variation of the action~\eqref{eq:matter} leads to an expression for the variation of the free energy in terms of the variations of the sources. In this case, the result is simply
\be
 \delta F = -\frac{2N_f\ell^3}{\kappa^2N_c} \sigma \delta m_q \ .
\ee
Choosing a normalization where the free energy vanishes when $\hat m_q = 0 = \hat \sigma$, the spiral equations Eq.~(\ref{mqh}) and Eq.~(\ref{sigmah}) with Eq.~(\ref{alphah}) imply that
\bea
 \hat F &=&-\frac{1}{16} \left(\frac{r_\uv}{r_\ir}\right)^4 \left[1-\left(1+\nu ^2\right) \cos  \left(2\nu \log \frac{r_\ir}{r_\uv}+2\arctan\frac{\nu B}{A+2B} \right)\right]\\
 &=& -\frac{1}{2}\hat m_q \hat \sigma -\frac{\nu^2}{16}\left(\frac{r_\uv}{r_\ir}\right)^4 \label{eq:freeencomp}
\eea
where $\hat F$ was normalized such that $\delta \hat F = - \hat \sigma \delta \hat m_q$.
From the first expression we see that $\hat F$ is non-monotonic along the spiral: the expression even changes sign as the ratio $r_\uv/r_\ir$ varies. This is expected because at some point $\delta \hat m_q$ must change sign on any spiral. However since $\nu$ is small, the value of $\hat F$ is typically negative. Moreover, since the overall scaling factor $r_\uv^4/r_\ir^4$ dominates the parameter dependence along the spiral, the lowest free energies are generally found for solutions with large $r_\uv/r_\ir$, i.e., least amount of scaling. This is seen clearly for the zero mass solutions from the second expression~\eqref{eq:freeencomp}: the first term vanishes and the second term is monotonic in the ratio.

Notice the result for the free energy also implies that Miransky scaling is effectively the upper limit for the amount of scaling of stable solutions. From Fig.~\ref{fig:spirallog} we see that for solutions with larger scaling, i.e., solutions below the green band, another solution with the same quark mass exists in or above the green band and therefore has lower free energy.

\subsection{PNGB dilaton mass}\label{sec:mass_toy}

Let us then discuss what the results for the dilaton mass from Sec.~\ref{sec:dilaton_mass} imply in the case of the toy model. For the toy model, the phase $\alpha$ of the background flavor field $X(r)$ is determined by Eq. (\ref{alpha_toy}). For zero quark mass $m_q=0$, $\alpha \simeq \nu$ and assuming further $\beta \gg \alpha$ and ${\cal A}=0$ (Neunmann IR B.C.) in Eq. (\ref{lmass}) as required for a light mode while taking $c_\uv \simeq 0$ from the UV normalizability,
\dis{ \label{eq:mass_small_alpha}
\omega^2 r_\ir^2 \simeq \frac{3\nu^2}{\sin^2 \beta}
} 
yielding a small mass $\omega \simeq \sqrt{3} \nu r_\ir^{-1}/\sin\beta$.

On the other hand, as discussed above, many solutions with scaling that differs from the standard Miransky law (i.e. $\beta \sim 1$) fail to satisfy $\alpha \sim {\cal O}(\nu)$ for non-zero quark mass $m_q \neq 0$, but $\alpha$ takes a larger value. For such solutions, imposing the Neunmann IR B.C. ${\cal A}=0$ again, we find 
\dis{ \label{eq:mass_large_alpha}
\omega^2 r_\ir^2 \simeq \frac{2\nu\sin\alpha }{\sin(\beta-\alpha)\sin \beta} \
}
so that the mass is still small for $\alpha \sim \beta \sim {\cal O}(1)$, yielding $\omega \sim \sqrt{\nu} \,r_\ir^{-1}$. Finally, in the case that the condensation is negligible compared with the quark mass $\sigma r_\uv^3 \ll m_q r_\uv$, Eq. (\ref{alpha_toy}) tells us $\alpha \simeq -\nu$, for which Neunmann IR B.C. ${\cal A}=0$ leads to a tachyon with 
\dis{
\omega^2 r_\ir^2 \simeq -\frac{\nu^2}{\sin^2\beta},
}
which indicates that the dilaton is unstable, and the Neunmann IR B.C. is not a correct condition for a vacuum.  
Therefore, if $\sigma r_\uv^3 \ll m_q r_\uv$, a light mode does not exist. 

\subsection{Two-point correlation function and the PNGB dilaton decay constant}

The two-point correlation function of the trace of the energy-momentum tensor allows us to compute the PNGB dilaton decay constant $f_D$ given the PNGB dilaton mass $m_D$. This is from the Partially Conserved Dilatation Current (PCDC) relation, which can be stated as \cite{Choi:2012kx,Kawaguchi:2020kce}
\bea
\lim_{q^\mu \rightarrow 0} \int d^4 x e^{iq\cdot x} \langle 0 | T \theta^\mu_\mu(x) \theta^\nu_\nu(0) | 0 \rangle = -if_D^2 m_D^2 \label{pcdc}
\eea
where $q_\mu \rightarrow 0$ but $q^2 \gg m_D^2 \approx 0$. Here we can find an explicit expression for the two-point function from the holographic principle in the case of the toy model. 
Starting from the action for the gauge-invariant fluctuation $\xi$ (see Appendix~\ref{app:fluctuations})
\dis{
{\cal S} = -\frac{\cal N}{2\kappa^2} \int d^4x dr \sqrt{-\det \bar g}  \left(\frac{\bar{X}'}{A'}\right)^2 {\bar g}^{M N} \partial_ M \xi \partial_N \xi ,
}
we get the following 4D generating functional by integrating by parts and applying the equation of motion for $\xi$:
\dis{
W[\xi] = \frac{\cal N}{2\kappa^2} \int d^4x \left. e^{3A(r)} \left(\frac{\bar{X}'}{A'}\right)^2 \xi \xi' \right|_{r=r_{\rm UV}}.
}
The holographic principle tells us that the two point function can be obtained as
\bea
\langle {\theta^{\mu}_\mu}(x) {\theta^{\nu}_\nu}(0) \rangle &=& -i \left.\frac{\delta^2 W}{\delta \psi(x) \delta \psi(0)}  \right|_{\psi=0}\,, \label{2pt}
\eea
where $\psi$ is defined as Eq. (\ref{psidef}) in Appendix \ref{app:fluctuations} which couples to the trace of the energy-momentum tensor. It
is related to the fluctuation $\xi$ as (Eq. (\ref{giv}) in Appendix \ref{app:fluctuations}), 
\dis{
\xi(x,r) = \psi(x,r) - \frac{A'}{\bar{X}'} \chi(x,r)
}
We may write the solution of the equation of motion for $\xi(x,r)$ in Eq. (\ref{eq:Besselsol}) as 
\dis{
\xi(x, r) = \frac{\eta(x, r) }{2\sin\left( \nu \ln \frac{r}{r_\uv}\right)+ \nu \cos\left( \nu \ln \frac{r}{r_\uv}\right)}
}
with
\dis{
\eta(x^\mu, r) = \sum_\omega \left[\frac{\textrm{Re}[J_{i \nu} (\omega r)]}{\textrm{Re}[J_{i \nu} (\omega r_{\rm UV})]} + c_2(\omega) \left( \textrm{Re}[Y_{i \nu} (\omega r)] + c_1(\omega) \textrm{Re}[J_{i \nu} (\omega r)] \right) \right] \eta_\omega(x^\mu)
}
where $\eta_\omega(x^\mu)$ is the 4D mass eigenstate with mass $\omega$. Here we take $\textrm{Re}[J_{i \nu} (\omega r)]$ as a source term while the next terms are taken to be VEV. 
Applying the UV boundary condition Eq. (\ref{xiuvbc}) for the VEV and the IR boundary condition Eq. (\ref{IRbc}) for the entire $\xi(x, r)$, we can determine $c_1(\omega)$ and $c_2(\omega)$. It turns out that
\dis{
c_2 (\omega) \simeq -\frac{\pi \nu}{2\cos \beta (\tan\beta+c_{\rm UV})} \frac{ \omega^2- m_D^2  - (2\nu/\tan \beta) r_\ir^{-2} }{\omega^2 - m_D^2} \label{c2q}
}
around the dilaton mass $\omega^2 \approx m_D^2$, where $c_{\rm UV}$ is defined as Eq. (\ref{cuv}) parametrizing the UV boundary condition, and we assume $ r_\ir^2 m_D^2 \ll 1$.

Finally the two point correlation function from Eq. (\ref{2pt}) comes out as
\dis{
\int d^4 x e^{iq\cdot x} \langle 0 | T \theta^\mu_\mu(x) \theta^\nu_\nu(0) | 0 \rangle &\simeq i \frac{{\cal N}\ell^3}{2\kappa^2} \frac{X_0^2}{r_\uv^2} \nu \left( 2\tan \beta + \frac{2}{\pi \nu} c_2(q) \left[ \cos \beta + \sin \beta (\tan \beta + c_{\rm UV})\right]\right) \label{2ptr}
} 
where $X_0$ is defined as Eq. (\ref{eq:Xwalkingsol}).
Therefore, we obtain a pole at $q^2 = m_D^2$ in the two point function from $c_2(q)$ in Eq. (\ref{c2q}).
Taking the limit $q_\mu \rightarrow 0$ but $q^2 \gg m_D^2 \approx 0$, and assuming $c_{\rm UV} \ll \tan \beta$, 
\bea
\lim_{q^\mu \rightarrow 0} \int d^4 x e^{iq\cdot x} \langle 0 | T \theta^\mu_\mu(x) \theta^\nu_\nu(0) | 0 \rangle 
\simeq -i  \frac{{\cal N}\ell^3}{2\kappa^2} \frac{X_0^2}{r_\uv^2}   \frac{2\nu}{\sin 2\beta}\,.
\eea

The PCDC relation Eq. (\ref{pcdc}) therefore tells us that
\bea
f_D^2 m_D^2 &\simeq& \frac{{\cal N}\ell^3}{2\kappa^2} \frac{X_0^2}{r_\uv^2}  \frac{2\nu}{\sin 2\beta} \,, \\
&=& \frac{{\cal N}\ell^3}{2\kappa^2} X^2(r_\ir) r_\ir^{-4}  \left(  \frac{2\nu}{\sin 2 \beta \sin^2(\beta-\alpha)} \right) \sim r_\ir^{-4}  \left(  \frac{\nu}{\sin \beta \cos\beta \sin^2(\beta-\alpha)} \right). \nonumber \\
\eea
In order to satisfy the relation $f_D^2 m_D^2 \sim r_\ir^{-4}$ as argued by Miransky and Gusynin \cite{Miransky:1989qc}, therefore, we need $\beta -\alpha \sim \sqrt{\nu}$ while $\beta \sim {\cal O}(1)$ as shown in \cite{CruzRojas:2023jhw}, or $\beta \sim \nu^{1/3} > \alpha$. This is consistent with a parametrically light dilaton which requires $\beta \gg \nu$ as we have shown in Section \ref{sec:dilaton_mass}.


\section{UV completion with flowing coupling constant} \label{sec:running} 

Let us then study a more general case in which, instead of the choice $V(\phi)=\Lambda_5$ of the toy model, consider
a more general dilaton potential $V(\phi)$. 
The potential then gives rise to solutions of the dilaton, which depend on the holographic coordinate $r$. This ``holographic RG flow'' of the dilaton corresponds to running of the coupling on the field theory side. The inclusion of the dilaton replaces the direct gluing of the near-boundary solutions ($r<r_\uv$) to the solutions in the walking regime ($r_\uv<r<r_\ir$), which we did in the toy model of the previous section, by a smooth but non-trivial flow.

In our setup, where we want to study the physics near the conformal transition to QCD, a complete theory is expected to include both an UV and an IR fixed points, which are realized through extrema of the dilaton potential. A possible choice for the UV fixed point (chosen in the IHQCD and V-QCD models~\cite{Gursoy:2007cb,Gursoy:2007er,Jarvinen:2011qe}) is such that the UV fixed point is at $\phi = -\infty$ and the logarithmic UV running of the dilaton is driven by corrections to the potential that are suppressed by powers of $e^{\phi}$. A simpler choice is to set the UV fixed point at a finite value, so that the mass of the dilaton around the fixed point controls the dimension $\Delta_g$ of the dual operator $\mathrm{Tr}G^2$ (see, e.g.,~\cite{Gubser:2008ny}). For stable solutions with positive mass, the dimension satisfies $2<\Delta_g<4$. Note that the case $\Delta_g =4$ is physically motivated as it is the weak-coupling value of the dimension of the operator -- this choice is obtained in the IHQCD model or by considering higher-order perturbations of the potential around the fixed point instead of a mass term  (see Appendix~\ref{app:RG_flow}).
For our purposes the precise value of $\Delta_g$ is, however, not important, as the physics we are interested in here mostly arises from the behavior of the solutions near the IR fixed point. Therefore we mostly consider the simple case with fixed point at a  finite value of $\phi$ and non-zero mass, keeping $\Delta_g$ as a free parameter.
Let us denote the fixed point values of the dilaton as $\phi_\mathrm{UV}$ and $\phi_\mathrm{IR}$.

We will first define the model explicitly. We then discuss how this more general setup is related to the toy model in the previous section. As we shall show, the simple UV completion of the toy model works as a characteristic example of the general case. For example, the phases of the tachyon oscillation, and how they depend on $\nu$, is qualitatively similar in the toy model as for general UV completions with non-trivial dilaton flow.

A complete model with proper gravity sector, however, allows us to extend the analysis in other directions. 
In particular, we can study the flavor contributions to the one- and two-point functions of the Yang-Mills sector, and carry out a more precise analysis of the presence of light scalar states.

\subsection{A concrete model with gravity and matter sectors} 

For a simple concrete model that is able to realize the full picture (see Sec.~\ref{sec:RGflow}), let us consider\footnote{Full discussion of the boundary terms and counterterms is given in the Appendix.}
\be \label{eq:gravSc}
 S_\mathrm{grav} = \frac{1}{2\kappa^2} \int d^5x \sqrt{-\det g}\left[R - g^{MN}\partial_M \phi \partial_N \phi + V(\phi) \right] \ ,
\ee
where 
the potential will have extrema both for the UV and IR fixed point values of $\phi$, and
\be \label{eq:Xaction}
 S_\mathrm{matter} = -\frac{\mathcal{N}}{2\kappa^2}\int d^5x \sqrt{-\det g}\ \frac{1}{N_f} \mathrm{Tr}\left[g^{MN}\partial_M X^\dagger \partial_N X - m_X(\phi)^2 X^\dagger X \right]
\ee
where we allowed the scalar mass to depend on $\phi$. While this is not the most general action, it is enough to capture all possibilities in the regions that we are interested in, leading to largely model independent results, as we argue below. 
The normalization constant $\mathcal{N}$ could be eliminated by rescaling $X$, but we choose to keep it so that we can keep track of terms that are small in the probe limit $S_\mathrm{matter} \ll S_\mathrm{grav}$. However, we stress that it is not necessary to take $\mathcal{N}$ to be small for the analysis of this section, because $X$ turns out to be suppressed in the regimes we study in the following, so that the probe approximation is obtained dynamically.

As above, we use the ``conformal coordinates'', i.e., we fix the freedom of diffeomorphism transformations such that
\be
 ds^2 = e^{2A(r)}\left(dr^2-dt^2+d\mathbf{x}^2\right)
\ee 
where $r$ is the bulk coordinate. The equations of motion (see Appendix~\ref{app:eoms}) include the Einstein equations, which also include the terms arising from the flavor action~\eqref{eq:Xaction}, and the dilaton equation, but the only equation that is needed in the near boundary analysis is the equation for $X$:
\be \label{eq:Xbgeq}
 \partial_r\left(e^{3A(r)} X'(r)\right) + e^{5A(r)}m_X(\phi(r))^2 X = 0 \ .
\ee

In order to implement the picture with the RG flow discussed above in Sec.~\ref{sec:RGflow}, we choose the potential $V$ such that it has a minimum at some value $\phi_\uv$ of the dilaton that we can set to zero without loss of generality, and a maximum at $\phi_\ir$. We further assume that the potential is regular enough so that a flow solution between the fixed points exists. The geometry is then asymptotically AdS$_5$ both in the UV and in the IR (so long as $X$ is so small that its effect on the geometry is negligible) with the AdS radii given by
\be
 V(0) = \frac{12}{\ell^2} \ , \qquad V(\phi_\ir) = \frac{12}{\ell_*^2}\ ,
\ee
respectively. The values of $m_X$ at the fixed points are then linked to the UV and IR dimensions via
\be \label{eq:FPmasses}
 m_X(0)^2 = \frac{\Delta_\uv(4-\Delta_\uv)}{\ell^2}\ , \qquad m_X(\phi_\ir)^2 = \frac{\Delta_\ir(4-\Delta_\ir)}{\ell_*^2} \ .
\ee
For simplicity, we will assume below that $\Delta_\uv$ takes the free field value for the $\bar q q$ operator, $\Delta_\uv =3$. We are mostly interested in the (walking) case where the IR dimension slightly violates the BF bound, i.e., $\Delta_\ir = 2 +i \nu$ with $\nu \ll 1$.

\subsection{Solutions with scale separation and linking to the toy model} \label{sec:linktoy}

We will now argue that the results obtained by the simple toy model of Sec.~\ref{sec:toymodel} continue to hold at the qualitative level for the more general action of Eqs.~\eqref{eq:gravSc} and~\eqref{eq:Xaction}. We first discuss the results for the background and the Efimov spiral and then the fluctuations. Some assumptions are however needed: the general discussion below will only be valid in the presence of scaling, i.e., we must have $r_\uv \ll r_\ir$ in the limit $\nu \to 0$. Recall that in Sec.~\ref{sec:toymodel} we found that depending on the choice of model in the IR, the separation of UV and IR scales may also be absent. In such cases we cannot derive model independent results by using the techniques presented here.

\subsubsection{Scales of the flow and the geometry}

Let us first discuss the general structure of the background solution, which was already sketched in Sec.~\ref{sec:setup}. The idea is that the solution is divided into two parts which become well separated in the limit $\nu \to 0$:
\begin{enumerate}
    \item For $r\sim r_\uv$, the background flows from the (trivial) UV fixed point to the IR fixed point.
    \item For $r \sim r_\ir$, the background flows from the IR fixed point to the (unspecified) regular IR solution.
\end{enumerate}
The two parts of the background solution may be formally obtained from a generic background solution at finite positive $\nu$ by taking the limit $\nu \to 0$ keeping either $r_\uv$ or $r_\ir$ fixed (see Sec. 5.3 and Appendix~I in~\cite{Arean:2013tja}). Near both fixed points, the geometry is close to AdS$_5$:
\begin{align} \label{eq:bgUVass}
 e^{A(r)} &\approx \frac{\ell}{r} \ ,& \qquad \phi(r) &\approx \phi_\uv \ ,& \qquad &(r \ll r_\uv)&\\ \label{eq:bgwalking}
 e^{A(r)} &\approx \frac{\ell_*}{r} \ ,& \qquad \phi(r) &\approx \phi_\ir \ ,& \qquad &(r_\uv \ll r \ll r_\ir) \ .&
\end{align}

In order to make these statements precise, let us then give explicit definitions for the UV and IR scales $r_\uv$ and $r_\ir$.
First, $r_\uv$ corresponds to UV scale around which $\phi$ starts to deviate from $\phi_\uv$ significantly, while around $r_\ir$ the field $X$ grows large so that its backreaction and the non-linearities in the action for $X$ will become important.  To make this explicit, we consider potential of the form
\be
\label{eq:VphiUVtext}
 V(\phi) = \frac{12}{\ell^2} + \frac{\Delta_g(4-\Delta_g)}{\ell^2}(\phi-\phi_\uv)^2 + \morder{(\phi-\phi_\uv)^4}
\ee
where $0<\Delta_g<4$. Since the perturbative dimension of the dual operator, $G^2$ is four, it is natural to take $\Delta_g$ to be close to this value. The boundary expansion for the dilaton becomes
\be \label{eq:phisphiv}
 \phi(r) = \phi_\uv+ \phi_s r^{4-\Delta_g} + \phi_v r^{\Delta_g} + \cdots
\ee
We may now define $r_\uv$ as 
\be \label{eq:rUVdef}
 r_\uv = \frac{1}{\phi_s^{1/(4-\Delta_g)}}
\ee
so that the source term becomes
\be
 \phi(r) \approx \phi_\uv+ \left(\frac{r}{r_\uv}\right)^{4-\Delta_g} 
\ee
as $r \to 0$. On the other hand, $r_\ir$ may be given by
\be \label{eq:rIRdef}
 X(r_\ir) = \widehat X
\ee
where $\widehat X$ is a fixed $\mathcal{O}(1)$ number.

\subsubsection{The flavor field and the Efimov spiral}

We start by discussing the behavior near the boundary, $r \sim r_\uv$. The first observation about the background is that the equation of motion for $X$ is linear. In our case this appears to be the case by default, since the action~\eqref{eq:Xaction} is quadratic in $X$, but this is expected to hold in general: $X$ is small for $r \ll r_\ir$, and the non-linear terms are negligible. Actually, we will find $X(r\sim r_\uv) \propto (r_\uv/r_\ir)^2$ which reflects the IR dimension $\Delta_\ir$ for $\nu \to 0$. The non-linear terms will only become important in the IR, $r\sim r_\ir$, and in part determine the IR dynamics which we parametrize in this work through boundary conditions. Moreover notice that even though the action~\eqref{eq:Xaction} is linear in $X$, non-linearities could still arise due to the backreaction to the geometry. Therefore it is important even in the current case that $X$ is suppressed in the region that we are analysing.

Due to linearity and the smallness of $X$, the solution to~\eqref{eq:Xbgeq} can be written as the sum of the source and vev terms:\footnote{For generic actions the source term is only well-defined up to a multiple of the vev term, see, e.g.~\cite{Jarvinen:2015ofa}. This ambiguity will not affect the analysis in this article.}
\be \label{eq:Xcompdefs}
 X(r) = m_q r_\uv X_1(r) + \sigma r_\uv^3 X_2(r) \ , \qquad (r \ll r_\ir)
\ee
where we included powers of $r_\uv$ in order to make the basis functions $X_i$ dimensionless.
Therefore the components satisfy the UV asymptotics
\be \label{eq:Xcompass}
 X_1(r) =\frac{r}{r_\uv} \left[1+\morder{r^2}\right] \ , \qquad  X_2(r) =\left(\frac{r}{r_\uv}\right)^3\left[1+\morder{r^2}\right]\ , \qquad (r \ll r_\uv)
\ee
where we inserted $\Delta_\uv=3$. The power in the subleading corrections may depend on the details of the action, the typical value of 2 is assumed here. When $r_\uv \ll r \ll r_\ir$ (which is obtained by taking $\nu$ to be small), the solutions take the form given in~\eqref{eq:Xwalkingsol}:
\begin{align} \label{eq:Xwalkingcomps}
X_i(r) &\approx C_i^{(X)} \left(\frac{r}{r_\uv}\right)^2\sin\left(\nu \log \frac{r}{r_\uv}+ \rho_i\right)& \nonumber\\
 &= C_i^{(X)} \cos\rho_i\left(\frac{r}{r_\uv}\right)^2\sin\left(\nu \log \frac{r}{r_\uv}\right) & \\\nonumber 
 &\phantom{=} + C_i^{(X)} \sin\rho_i\left(\frac{r}{r_\uv}\right)^2\cos\left(\nu \log \frac{r}{r_\uv}\right)\ , \qquad (r_\uv \ll r \ll r_\ir) &
\end{align}
where $i=1,2$. That is, the flow of the solution for $r\sim r_\uv$ implies a non-trivial transformation between the bases that are natural near the boundary ($r \ll r_\uv$) and the walking regime $r_\uv \ll r \ll r_\ir$. The coefficients $C_i^{(X)} \cos\rho_i$ and $C_i^{(X)} \sin\rho_i$ form a transformation matrix between the bases which is non-trivial due to the flow. Note that both~\eqref{eq:Xcompass} and~\eqref{eq:Xwalkingcomps} are valid in the region ($r \ll r_\ir$) where $X$ is small and non-linear terms in its equation of motion can be safely neglected, as we have done here. This guarantees that the transformation is linear. Moreover, the definition of the coefficients $C_i^{(X)}$ and $\rho_i$ makes no reference to the values of $m_q$ and $\sigma$, so the coefficients are independent of these parameters.

At this point we can compare to the toy model of Sec.~\ref{sec:toymodel}: for this model, explicit results for the coefficients can be read off from Eq.~(\ref{alpha_toy}) and Eq.~(\ref{X0_toy}):
\begin{align}
\begin{aligned} \label{eq:toyCalphavals}
 \rho_1 &= -\arctan\nu \ , \qquad \qquad \rho_2 = +\arctan\nu \ , &  \\
 -C_1^{(X)} &= 
 C_2^{(X)} = \frac{1}{\nu}\sqrt{1+\nu^2} \approx \frac{1}{\nu} \ . &
\end{aligned} \qquad \textrm{(toy model)}
\end{align}
However for general model with unspecified form of the dilaton potential, the coefficients cannot be determined analytically. We do expect that they typically behave in the limit $\nu \to 0$ in the same way as in the toy model; we will discuss this below.

Let us then discuss the solution at $r\sim r_\ir$. Since we do not specify the model in the IR, we can only parametrize the asymptotic behavior of $X$ for $r \ll r_\ir$ where the background flows close to the fixed point:
\be \label{eq:XIRgen}
 X(r) = X_\ir \left(\frac{r}{r_\ir}\right)^2 \sin\left(\nu \log \frac{r}{r_\ir}-k_\ir\right) \ , \qquad (r_\uv \ll r \ll r_\ir) \ .
\ee
The point here is that as the background is expected to approach a well defined flow solution between the IR fixed point and some regular IR behavior as $\nu \to 0$ with $r_\ir$ fixed, we expect that $X_\ir$ and $k_\ir$ take values which are independent of the UV boundary conditions (i.e., the value of $m_q$) as well as the UV scale $r_\uv$ as $\nu \to 0$, and solely determined by IR regularity (see also Appendix~\ref{app:bcs}). We also remark that the solution~\eqref{eq:XIRgen} naively appears to depend on our somewhat arbitrary choice of the IR scale in~\eqref{eq:rIRdef}. For a fixed IR model, such dependence however should be absent. To make this explicit, let us pick another scale $r_\ir'$ defined by $X(r_\ir') = \widehat X'$. Let us denote the new IR parameters defined by using $r_\ir'$ instead of $r_\ir$ in~\eqref{eq:XIRgen} by $X_\ir'$ and $k_\ir'$. For the functional form of the solution in~\eqref{eq:XIRgen} to remain the same, we must have that
\be \label{eq:IRinvar}
 \frac{X_\ir}{r_\ir^2} = \frac{X_\ir'}{(r_\ir')^2} \ , \qquad k_\ir = k_\ir' - \nu \log \frac{r_\ir}{r_\ir'} \ .
\ee
In other words, the combinations $X_\ir/r_\ir^2$ and $k_\ir +\nu \log r_\ir$ are independent of the definition of the IR scale.

Comparing to Sec.~\ref{sec:toymodel}, we see that in the toy model (where we also have a single natural choice of $r_\ir$ as the cutoff scale)
\be \label{eq:kIRtoy}
 \tan k_\ir = \frac{\nu B}{A+2B} \ ,  \qquad \textrm{(toy model)}
\ee
where the parameters $A$ and $B$ determined the boundary condition in~\eqref{eq:IRbcgen}. The value of $X_\ir$ was left undetermined in the toy model as it did not affect the results. Again, for generic model the values of $k_\ir$ and $X_\ir$ cannot be analytically determined. Nevertheless, as in the toy model, we expect that generic models for the IR behavior will give phases that are small, i.e., $k_\ir \sim \nu$ but some special models may lead to larger phase factors.

Now similarly as in the toy model requiring that the solution in~\eqref{eq:Xwalkingcomps} is the same as~\eqref{eq:XIRgen} leads to the spiral equations for the quark mass and the condensate:
\begin{align} \label{eq:mqspiralgen}
m_q r_\uv &= -\frac{X_\ir}{C_1^{(X)}}\left(\frac{r_\uv}{r_\ir}\right)^2 \frac{\sin\left(\nu\log \frac{r_\ir}{r_\uv}+k_\ir+\rho_2\right)}{\sin\left(\rho_1-\rho_2\right)} & \\ \label{eq:sigmaspiralgen}
\sigma r_\uv^3 &= \frac{X_\ir}{C_2^{(X)}}\left(\frac{r_\uv}{r_\ir}\right)^2 \frac{\sin\left(\nu\log \frac{r_\ir}{r_\uv}+k_\ir+\rho_1\right)}{\sin\left(\rho_1-\rho_2\right)} \ .&
\end{align}
Since the result has the same form as in the previous section, see Eq.~(\ref{mqh}), Eq.~(\ref{sigmah})  and~\eqref{alphah}, it is immediate that the structure is similar as in the toy model. That is, a the theory with RG flow near the boundary will show the structure depicted in Figs.~\ref{fig:spirallinear} and ~\ref{fig:spirallog}. In particular, for massless solutions we find that
\be \label{eq:zeromassscaling}
 \frac{r_\uv}{r_\ir} = \exp\left(-\frac{\pi n- k_\ir-\rho_2}{\nu}\right) \ , \qquad (m_q = 0) \ ,
\ee
where $n$ is an integer. If $k_\ir+ \rho_2$ is small $\sim \nu$ and positive, which is true for typical boundary conditions in the toy model, $n=1$ is a Miransky scaling solution that is likely\footnote{Even if $k_\ir+ \rho_2>0$, we cannot strictly exclude solutions that do not show any scaling at all because our approach does not cover such solutions -- scale separation between the UV and the IR is required to obtain all our results.} to be the vacuum of the theory. We will discuss this point in a bit more detail in the next subsection.

\subsubsection{Additional constraints}

An additional exact result for the RG flow can be obtained by considering the Wronskian for $X$, i.e.,
\be \label{eq:Wdef}
 W(r) = X_1'(r) X_2(r) -X_1(r) X_2'(r) 
\ee
where $X_1$ and $X_2$ could be any solutions to~\eqref{eq:Xbgeq}, but without loss of generality we can take them to be the solutions defined through~\eqref{eq:Xcompdefs}. Assuming that, thanks to the suppressed backreaction of the $X$ to the geometry for $r \ll r_\ir$, the dilaton flow is the same for both solutions, we immediately obtain for the Wronskian, using the equation of motion for $X$,
\be \label{eq:Wsol}
 W(r) = C_W e^{-3A(r)} \ , \qquad (r \ll r_\ir)\ .
\ee
Near the boundary, inserting the expressions from~\eqref{eq:bgUVass} and from~\eqref{eq:Xcompass} in~\eqref{eq:Wsol} and in~\eqref{eq:Wdef}, respectively, we find that
\be
 W(r) \approx C_W \frac{r^3}{\ell^3} \approx -2\frac{r^3}{r_\uv^4} \ .
\ee
Therefore
\be
 C_W = -2\frac{\ell^{3}}{r_\uv^4} \ .
\ee
In the same way using the expressions near the IR fixed point from~\eqref{eq:bgwalking} and from~\eqref{eq:Xwalkingcomps}, we obtain
\be
 W(r) \approx C_W \frac{r^3}{\ell_*^3} \approx -\frac{\nu  r^3 C_1^{(X)} C_2^{(X)}  \sin \left(\rho_1-\rho_2\right)}{r_\uv^4} \ .
\ee
We then note that~\eqref{eq:Wsol} holds both near the boundary and in the walking regime, and in particular the coefficient $C_W$ has the same value in both regions. Equating the two expressions for this constant gives
\be \label{eq:Widentity}
- r_\uv^4 C_W = 2 \ell^3 =  \nu  C_1^{(X)} C_2^{(X)}  \sin \left(\rho_1-\rho_2\right) \ell_*^3\ ,
\ee
where the latter equality is an additional exact constraint for the coefficients $C_i^{(X)}$ and $\rho_i$. As a check, the expressions for the toy model from~\eqref{eq:toyCalphavals} satisfy this relation (even without imposing the limit $\nu \to 0$) if we set $\ell_*=\ell$ (as required by the absence of the holographic RG flow in the toy model).

The identity~\eqref{eq:Widentity} may look like an academic rather than a useful result. However, as it turns out, it does have significant physical implications. In particular, it fixes the handedness of the spiral in~\eqref{eq:mqspiralgen}. This can be seen from the analysis of the free energy that we discuss below, but also directly from the spiral as follows. Note that we can apply a linear transformation to the mass as the condensate to present the spiral as
\begin{align} \label{eq:mqtrans}
\tilde m_q r_\uv &\equiv m_q r_\uv + \cos(\rho_2-\rho_1)\frac{C_2^{(X)}}{C_1^{(X)}}\sigma r_\uv^3 &\nonumber \\
&= -\frac{X_\ir}{C_1^{(X)}}\left(\frac{r_\uv}{r_\ir}\right)^2 \cos\left(\nu\log \frac{r_\ir}{r_\uv}+k_\ir+\rho_1\right) &\\
\sigma r_\uv^3 &= \frac{X_\ir}{C_2^{(X)}}\left(\frac{r_\uv}{r_\ir}\right)^2 \frac{\sin\left(\nu\log \frac{r_\ir}{r_\uv}+k_\ir+\rho_1\right)}{\sin\left(\rho_1-\rho_2\right)} & \label{eq:sigmatrans}
\end{align}
where the transformation is defined by the first line of~\eqref{eq:mqtrans}.
This linear transformation of $m_q$ to $\tilde m_q$ does not affect the handedness; it corresponds to a rotation of the $\sigma$ axis on the $(m_q,\sigma)$ plane (with rotation angle less than $\pi/2$). As one sees from the coefficients in~\eqref{eq:mqtrans} and~\eqref{eq:sigmatrans}, the handedness of the resulting spiral is characterized by the sign of the product $C_1^{(X)} C_2^{(X)}  \sin \left(\rho_1-\rho_2\right)$ and therefore fixed by the identity~\eqref{eq:Widentity}.

Finally, let us analyze the behavior of the coefficients $C_i^{(X)}$ and $\rho_i$ as $\nu\to 0$. This is important: for example, as we see from~\eqref{eq:mqspiralgen}, the amount of scaling for zero mass solutions crucially depends on the behavior of the coefficient $\rho_2$. The result from the toy model in~\eqref{eq:toyCalphavals}  suggests that typically $\rho_i \sim \nu$ and $C_i^{(X)}\sim 1/\nu$. However the relation~\eqref{eq:Widentity}, which is the only exact result we can use, is not enough to derive the scalings. We do notice that this identity depends on $\nu$ explicitly, so some of the coefficients must vanish or diverge as $\nu \to 0$. A way to obtain more (but less precise) information is to assume that $X_i$ are continuous near $r \sim r_\uv$. That is, the expressions in~\eqref{eq:Xcompass} and~\eqref{eq:Xwalkingcomps} must scale in the same way if we set $r = r_\uv$, unless there are accidental cancellations. This implies
\be
 C_i^{(X)} \sin\left(\morder{\nu}+\rho_i\right) \sim 1
\ee
as $\nu \to 0$. Now the most simple, symmetric solution to these relations and to~\eqref{eq:Widentity} is the same scaling, which is obtained for the toy model,
\be \label{eq:standardscaling}
 \rho_i \sim \nu \ , \qquad C_i^{(X)}\sim \frac{1}{\nu} \ ,
\ee
and we expect that this is indeed the behavior for generic choices of the potential $V(\phi)$. We also note that for generic backgrounds the natural expectation for the scaling of the IR phase is $k_\ir \sim \nu$: this is obtained in the toy model,~\eqref{eq:kIRtoy}, and is also consistent with the reparametrization relation~\eqref{eq:IRinvar}. Then indeed $k_\ir + \rho_2 \sim \nu$ so that~\eqref{eq:zeromassscaling} leads to the standard Miransky scaling, $r_\uv/r_\ir \sim \exp(-\pi n/\nu)$.

However, other solutions than~\eqref{eq:standardscaling} are possible, for example
\be
 \rho_1 \sim \nu \ , \quad \rho_2 \sim 1 \ , \qquad C_1^{(X)} \sim \frac{1}{\nu} \ , \quad C_2^{(X)} \sim 1 \ ,
\ee
which would lead to a non-Miransky scaling at zero quark mass as seen from~\eqref{eq:zeromassscaling}. 

\subsection{Fluctuations and light modes} \label{sec:flucts} 

Let us then discuss the fluctuations. 
We focus in the scalar, i.e., rotationally invariant fluctuation modes. To this end, we turn on the following fluctuations (see Appendix~\ref{app:fluctuations} for more details)
\be
 \phi \mapsto \phi + \varphi \ , \qquad X \mapsto X + \chi \ , \qquad g_{\mu\nu} \mapsto g_{\mu\nu} + 2\left(\eta_{\mu\nu}-\frac{\partial_\mu\partial_\nu}{\eta^{\alpha\beta}\partial_\alpha\partial_\beta}\right)\psi
\ee
where $g_{\mu\nu}$ are the space-time components of the metric, and $\varphi$, $\chi$ and $\psi$ are the infinitesimal fluctuations. As we will take the fluctuations to be plane waves in space-time coordinates with non-zero mass, the inverse of $\partial^2$ is well-defined here. To proceed, we can define the following diffeormorphism invariant combinations: 
\be \label{eq:fluctdefs}
 \zeta = \psi - \frac{A'}{\phi'} \varphi \ , \qquad \xi = \psi - \frac{A'}{X'} \chi \ .
\ee
For this model the fluctuation equations of the spin zero sector then become in the conformal coordinates:
\begin{align}
\label{eq:xieq}
 \xi''(r) + \partial_r \log \frac{e^{3 A(r)} X'(r)^2}{A'(r)^2}\ \xi'(r) 
 +\frac{e^{2A(r)}}{X'(r)^2}{\mathcal{M}}(r)(\xi(r)-\zeta(r)) +m^2\xi(r)
 &= 0& \\
\label{eq:zetaeq}
 \zeta''(r) + \partial_r \log \frac{e^{3 A(r)} \phi'(r)^2}{ A'(r)^2}\ \zeta'(r)
  +\frac{\mathcal{N}e^{2A(r)}}{\phi'(r)^2}{\mathcal{M}}(r)(\zeta(r)-\xi(r)) +m^2\zeta(r)
 &= 0&  
\end{align}
where we inserted a plane wave form $\sim \exp(-ik_\mu x^\mu)$ with $m^2 = k^2$ for the fluctuations, and the mass term ${\mathcal{M}}(r)$ is a complicated function of the background and is given explicitly in Appendix~\ref{app:fluctuations}.

The relevant field to study here is the gauge invariant combination $\xi$, which is related to the fluctuation of $X$ near the boundary. Near the boundary (see also Appendix~\ref{app:fluctuations}), the flavor action is suppressed, which means that the mass term in the equation for $\zeta$ in~\eqref{eq:zetaeq} is suppressed. Consequently, the fields are decoupled, so it is enough to study the equation
\be
 \xi''(r) + \partial_r \log \frac{e^{3 A(r)} X'(r)^2}{A'(r)^2}\ \xi'(r) 
 +\frac{e^{2 A(r)}}{X'(r)^2}\mathcal{M}(r)\xi(r) +\omega^2\xi(r)
 = 0
\ee
Interestingly, redefining here $\xi(r) = A'(r)\hat \xi(r)/X'(r)$, applying the equations of motion for the background fields, and consistently dropping terms $\propto \mathcal{N}$ which are suppressed near the boundary, we find that
\be \label{eq:uvflucts}
 \partial_r\left(e^{3A(r)} \hat \xi'(r)\right) + e^{5A(r)}m_X(\phi(r))^2 \hat \xi(r) +e^{3A(r)}\omega^2 \hat \xi(r)= 0 \ ,\qquad (r \ll r_\ir)\ .
\ee
This equation, apart from the frequency dependent term, agrees exactly with the background equation in~\eqref{eq:Xbgeq}. This is not surprising, as when the backreaction of the flavor to the background is suppressed, the flavor action is exactly quadratic which implies that the equations for the background and the fluctuations match.

Unless the mass of the fluctuation is extremely high, i.e., assuming that $\omega \ll 1/r_\uv$, the normalizable solution is therefore identified with $X_{2}$ solution for the background defined in~\eqref{eq:Xcompdefs}. When the mass is small, i.e., when $\omega \lesssim 1/r_\ir$, we can write
\be
 \hat \xi (r) \propto \left(\frac{r}{r_\uv}\right)^2 \sin \left(\nu\log\frac{r}{r_\uv}+\rho_2\right) \ , \qquad (r_\uv \ll r \ll r_\ir)
\ee
where $\rho_2$ is the same phase as in the background solution having zero quark mass. When the mass of the fluctuation $\omega$ grows larger, we can insert the fixed point solution from~\eqref{eq:bgwalking}, which leads to
\be
 \hat \xi''(r) - \frac{3}{r}\hat \xi'(r) + \frac{4+\nu^2}{r^2}\hat \xi(r) +\omega^2 \hat \xi(r)=0 \ , \qquad (r_\uv \ll r\ll r_\ir) \ .
\ee
Here we inserted the IR mass value for the flavor field $X$ from~\eqref{eq:FPmasses}. The solution is the same as in the toy model (see~\eqref{eq:Besselsol}),
\be \label{eq:hatxisol}
 \hat \xi(r) = C_1 r^2\ \mathrm{Re}\left[J_{i\nu}(\omega r)\right]+C_2 r^2\ \mathrm{Re}\left[Y_{i\nu}(\omega r)\right] \ .
\ee
Below we will use these solutions to analyze correlators and the masses of the scalar modes.

\subsubsection{Two-point correlation function of the quark bilinear and the UV completion}

Let us then analyze the two-point functions. We first focus on the two-point function of the quark bilinear $\bar q q$, which is the easiest correlator to analyze. Also, it is expected to show the same potentially light mode as the correlator of the trace of the energy-momentum tensor $\theta_\mu^\mu$. 
We also discuss how the result for the two-point functions can be expressed in terms of coefficients of the solutions in the walking regime. This allows us to interpret how the results of the UV-completed setup are linked to simpler models where the UV completion is neglected. 

We first write:
\be \label{eq:xiwalks}
 \hat \xi(r) = \frac{X'(r)}{A'(r)}\xi(r)
 = \Xi_+ \left(\frac{r}{r_\ir}\right)^{2+i\nu} +  \Xi_- \left(\frac{r}{r_\ir}\right)^{2-i\nu}
\ee
in the walking regime, $r_\uv \ll r \ll r_\ir$. 
Note that this expression makes no reference to the UV completion, therefore the coefficients $\Xi^{\pm}$ are  determined (up to an irrelevant normalization factor) solely in terms of IR boundary conditions.

In the notation of Section~\ref{sec:linktoy}, the exact source and vev include additional non-trivial phase factors $\rho_{i}$ (see, e.g.,~\eqref{eq:Xwalkingcomps}). The phase factors were defined for the background $X$, but as we argued above, the field $\hat \xi(r)$ satisfies the same equation as the background for $r \ll r_\ir$ so long as the frequency of the fluctuation is small, $\omega \sim 1/r_\ir$. Therefore, the (similarly defined) phase factors for the fluctuations exactly match with those of the background. To use this information, we may first rewrite~\eqref{eq:xiwalks} as
\begin{align}
   \hat \xi(r)&= \frac{e^{-i \rho_2} \left(\frac{r_\uv}{r_\ir}\right)^{i\nu}\Xi_++e^{i \rho_2}\left(\frac{r_\uv}{r_\ir}\right)^{-i\nu}\Xi_-}{2 i\sin\left(\rho_1-\rho_2\right)} \left(\frac{r_\uv}{r_\ir}\right)^2\times & \\
  &\ \ \ \times \left[\left(\frac{r}{r_\uv}\right)^{2+i\nu}e^{i\rho_1} -   \left(\frac{r}{r_\uv}\right)^{2-i\nu}e^{-i\rho_1}\right] - & \\
  & -\frac{e^{-i \rho_1} \left(\frac{r_\uv}{r_\ir}\right)^{i\nu}\Xi_++e^{i \rho_1}\left(\frac{r_\uv}{r_\ir}\right)^{-i\nu}\Xi_-}{2 i \sin\left(\rho_1-\rho_2\right)} \left(\frac{r_\uv}{r_\ir}\right)^2\times &\\
 & \ \ \ \times\left[\left(\frac{r}{r_\uv}\right)^{2+i\nu}e^{i\rho_2} -   \left(\frac{r}{r_\uv}\right)^{2-i\nu}e^{-i\rho_2}\right]
\end{align}
which holds for $r_\uv \ll r \ll r_\ir$ if $\omega \sim 1/r_\ir$. Here the first (second) term gives the source (vev) of the UV-completed setup. Therefore the exact correlator of the bilinear  $\bar q q$ reads in terms of the data $\Xi_\pm$ from the walking region
\be \label{eq:exactG}
   G(\omega^2) \sim \frac{\mathcal{N}}{\kappa^2}  \frac{C_1^{(X)}}{C_2^{(X)}} \frac{e^{-i \rho_1} \left(\frac{r_\uv}{r_\ir}\right)^{i\nu}\Xi_++e^{i \rho_1}\left(\frac{r_\uv}{r_\ir}\right)^{-i\nu}\Xi_-}{e^{-i \rho_2} \left(\frac{r_\uv}{r_\ir}\right)^{i\nu}\Xi_++e^{i \rho_2}\left(\frac{r_\uv}{r_\ir}\right)^{-i\nu}\Xi_-} \ .
\ee

We then compare the exact result to some simple approximations which do not use any information about the UV-completion but only the solutions in the walkig region.
The oscillating solution of~\eqref{eq:xiwalks} may also be written as
\be
 \hat \xi(r) = \left(\Xi_+ + \Xi_-\right)\left(\frac{r}{r_\ir}\right)^{2}\cos\left(\nu \log\frac{r}{r_\ir} \right) + i\nu (\Xi_+-\Xi_-) \left(\frac{r}{r_\ir}\right)^{2} \frac{\sin\left(\nu \log\frac{r}{r_\ir} \right)}{\nu}
\ee
which becomes in the limit of small $\nu$:
\be
   \hat \xi(r)= \left(\Xi_+ + \Xi_-\right)\left(\frac{r}{r_\ir}\right)^{2} + i\nu (\Xi_+-\Xi_-) \left(\frac{r}{r_\ir}\right)^{2} \log\frac{r}{r_\ir}\ ,  \qquad (\nu \to 0)\  .
\ee
Therefore, one possibility of identifying the source and the vev in the walking regime, up to a proportionality constant, is to take them to be  $i\nu(\Xi_+-\Xi_-)$ and $\Xi_+ + \Xi_-$, respectively. Consequently, the correlator reads
\be \label{eq:Gasympt}
  G_\mathrm{asympt}(\omega^2) \sim \frac{\mathcal{N}}{\kappa^2} \frac{\Xi_+ + \Xi_-}{i\nu(\Xi_+-\Xi_-)}
\ee
However, since it is expected that the system will have a separate UV scale, it may be obvious that source and vev definitions based solely on IR behavior are too simplistic. Indeed, the approximation~\eqref{eq:Gasympt} does not seem to reflect the phase structure of the exact result~\eqref{eq:exactG} -- recall that the natural expectation for the phases $\rho_i$, which is also supported by the model of Sec.~\ref{sec:toymodel}, is that they are small, $\rho_i \sim \nu$.

Without going to details of the flow in the UV, a way of introducing the UV scale in the correlators is to set a cutoff at $r =r_\uv$ and require that the vev term vanishes at the cutoff~\cite{CruzRojas:2023jhw}. A decomposition which achieves this is
\begin{align}\label{eq:cutoffcomp}
  \hat \xi(r) &= \left[\left(\frac{r_\uv}{r_\ir}\right)^{i\nu}\Xi_++\left(\frac{r_\uv}{r_\ir}\right)^{-i\nu}\Xi_-\right]\times\nonumber& \\
 & \times\frac{1}{2}\left[\left(\frac{r}{r_\ir}\right)^{2+i\nu}\left(\frac{r_\uv}{r_\ir}\right)^{-i\nu}+\left(\frac{r}{r_\ir}\right)^{2-i\nu}\left(\frac{r_\uv}{r_\ir}\right)^{i\nu}\right] + \nonumber &\\
& +i\left[\left(\frac{r_\uv}{r_\ir}\right)^{i\nu}\Xi_+-\left(\frac{r_\uv}{r_\ir}\right)^{-i\nu}\Xi_-\right]\times\nonumber& \\
 & \times\frac{1}{2i}\left[\left(\frac{r}{r_\ir}\right)^{2+i\nu}\left(\frac{r_\uv}{r_\ir}\right)^{-i\nu}-\left(\frac{r}{r_\ir}\right)^{2-i\nu}\left(\frac{r_\uv}{r_\ir}\right)^{i\nu}\right] 
 \ .&
\end{align}
From here one can read the correlator:
\be \label{eq:Gcutoff}
  G_\mathrm{cutoff}(\omega^2) \sim \frac{\mathcal{N}i}{\kappa^2} \frac{\left(\frac{r_\uv}{r_\ir}\right)^{i\nu}\Xi_+-\left(\frac{r_\uv}{r_\ir}\right)^{-i\nu}\Xi_-}{\left(\frac{r_\uv}{r_\ir}\right)^{i\nu}\Xi_++\left(\frac{r_\uv}{r_\ir}\right)^{-i\nu}\Xi_-}
\ee
Notice that requiring the vev term to vanish at the cutoff does not uniquely determine the source function in~\eqref{eq:cutoffcomp}. It does, however, determine the coefficient of the source up to a trivial constant. This is the most important piece in the correlator~\eqref{eq:Gcutoff} as it appears in the denominator and controls the poles of the correlator, i.e., the spectrum. Redefinitions of the source function only modify the correlator such that the pole structure is not altered.

The result appears to be quite different from~\eqref{eq:Gasympt} and in better agreement with the exact correlator~\eqref{eq:exactG}, as essentially only the phase factors $\rho_i$, which are expected to be small, are missing. That is,
the cutoff correlator in~\eqref{eq:Gcutoff} does give a reasonable model for the exact correlator, and the locations of the modes are the same for a vanishing vev phase, $\rho_2=0$. 

\subsubsection{Structure of the correlator and light modes}

We will then proceed to compute the coefficients $\Xi^\pm$ of~\eqref{eq:xiwalks} and consequently the correlator of~\eqref{eq:exactG} explicitly. Recall that the solution for $\hat \xi$ in the walking regime was already given in terms of Bessel functions in~\eqref{eq:hatxisol}. We parametrize the IR boundary conditions as above, extending the solution to $r=r_\ir$. This is outside the region of validity of the analytic solution, but as we argue in Appendix~\ref{app:bcs}, considering generic boundary conditions for this solution is representative for boundary conditions imposed by more general IR dynamics for small masses of the fluctuations. Following conventions given above, we write
\be \label{eq:IRbcgenqq}
 \mathcal{A}\, \xi(r_\ir) + \mathcal{B}\, r_\ir \xi'(r_\ir) = 0\ . 
\ee
for the field $\xi = A' \hat \xi/X'$, assuming the walking region solution given in~\eqref{eq:xiwalks}. 

It is then straightforward to solve the coefficients $\Xi^{\pm}$ by expanding the Bessel functions at small $r$. Imposing the condition~\eqref{eq:IRbcgen}, we find that
\begin{align} \label{eq:Xipmres}
 \Xi^{-}&=\left(\Xi^{+}\right)^* = & \nn\\
  &=    i C_\Xi e^{i \gamma_\nu } \left(\frac{\omega r_\ir }{2}\right)^{-i \nu } \bigg\{ \left[\mathcal A (2 s_{\beta-\alpha}-\nu  c_{\beta-\alpha})+\mathcal B \nu  (2-i \nu ) e^{i (\beta -\alpha )}\right]J_{i \nu }(\omega r_\ir )- &\nonumber\\
  &\qquad \qquad -\mathcal B  (2 s_{\beta-\alpha}-\nu  c_{\beta-\alpha})\omega r_\ir  J_{1+i \nu }(\omega r_\ir )\bigg\} & 
\end{align}
where $\gamma_\nu$ was defined in~\eqref{eq:gammanudef} and $C_\Xi$ is a real coefficient. It can be computed analytically but the expression is complicated and the coefficient is irrelevant for us because it cancels in the correlator. The expression for the correlator is then found by inserting~\eqref{eq:Xipmres} in~\eqref{eq:exactG}.

Let us then discuss the mass of the potential light mode, identified as the pole of the correlator. Setting the denominator in~\eqref{eq:exactG} to zero, using the expressions~\eqref{eq:Xipmres}, we find that the pole is located at
\be
\omega^2 r_\ir^2 \approx \frac{8  s_{\beta -\rho_2} (\nu  c_{\alpha -\beta}+2 s_ {\alpha -\beta })\mathcal{A}-8 \nu  (\nu  c_ {\alpha -\rho_2}+2 s_{\alpha -\rho_2})\mathcal{B}}{ \left[\nu  (s_{\alpha -2 \beta +\rho_2}+3 s_ {\alpha -\rho_2})+2 c_{\alpha -2 \beta +\rho_2}-2 c_{\alpha -\rho_2}\right]\mathcal{A}+4 s_ {\alpha -\beta} (\nu  c_{\beta -\rho_2}+2 s_{\beta -\rho_2})\mathcal{B}}
\ee
where we dropped terms suppressed by $\omega^2$ or $\nu^2$. This expression is not very illuminating, so let us consider some physically motivated special cases. These include Neumann boundary conditions, $\mathcal{A} = 0$, for which the expression becomes
\be
\omega^2 r_\ir^2 \approx \frac{2 \nu  (\nu  c_{\alpha -\rho_2}+2 s_{\alpha -\rho_2})}{\left(\nu  c_{\beta -\rho_2}+2 s_{\beta -\rho_2}\right)s_{\beta-\alpha}} \ .
\ee
As expected from the toy model, this expression is suppressed in the limit of small $\nu$. 

We may further consider the case of zero quark mass, which means that the near-boundary solution for the background is proportional to the normalizable fluctuation mode, i.e., $\alpha = \rho_2$. Then we find
\be
\omega^2 r_\ir^2 \approx  \frac{2 \nu ^2 }{\left(2 s_{\beta-\alpha}+\nu  c_{\beta-\alpha}\right)s_{\beta-\alpha}} \ .
\ee
In particular, unless $\beta-\alpha$ is small, $\omega \sim \nu$ in this case.
In general, even if the quark mass is not zero, we expect that $\alpha \sim \nu \sim \rho_2$. Assuming this scaling, to leading order in $\nu$,
\be \label{eq:mass_gen_scaled}
\omega^2 r_\ir^2 \approx (\nu +2 \alpha -2 \rho_2)\frac{\nu}{s_\beta^2} \ ,
\ee
so that again $\omega \sim \nu$, if $\beta \sim 1$. If $\beta \sim \nu$, the light mode disappears as we also saw in the toy model. 
It is also interesting to check the case of the toy model, $\tan \rho_2 = \nu$, which gives
\be
\omega^2 r_\ir^2 \approx \frac{2 \nu  (2 s_\alpha -\nu  c_\alpha )}{\left(2 s_\beta -\nu  c_\beta \right)s_{\beta-\alpha}} \ .
\ee
Now it is immediate that if we take the same limits as in Sec.~\ref{sec:mass_toy}, we reproduce the expression~\eqref{eq:mass_large_alpha}.

One can also analyze the structure of the full correlator given in~\eqref{eq:exactG}. Assuming for simplicity Neumann boundary conditions and $\alpha\sim \rho_1 \sim \rho_2 \sim \nu$ (while $\beta \sim 1$), we obtain
\be
 G(\omega) \sim \frac{\mathcal{N}}{\kappa^2}  \frac{C_1^{(X)}}{C_2^{(X)}}\frac{\omega^2 - m_D^2 + 2 \nu  \frac{\left(\rho_1-\rho_2\right)}{s_\beta^2r_\ir^2}}{\omega^2 - m_D^2 }
\ee
with $m_D^2$ given in~\eqref{eq:mass_gen_scaled}. Note that the decay constant of the mode therefore depends on the phases $\rho_i$.

\subsubsection{Correlator of the trace of the energy momentum tensor}

Let us then comment on the correlator of $\theta_\mu^\mu$. Similarly as the one-point function considered in Appendix~\ref{app:correlators}, the correlator has contributions both from the gluon and from the flavor sector. The contribution from the flavor sector is expected to have similar structure as the correlator of the quark bilinear discussed above. In particular, it contains poles due to the light mode if such a mode is present. However, it is tricky to write compact expressions for the general correlator such as~\eqref{eq:exactG} in this case. The main reason is that, computing the correlator of $\theta_\mu^\mu$ requires adding sources both for the gauge invariant fields $\xi$ and $\zeta$ in~\eqref{eq:xieq} and in~\eqref{eq:zetaeq} (see Appendices~\ref{app:fluctuations} and~\ref{app:correlators}). The mass term in the $\xi$-equation~\eqref{eq:xieq} cannot be neglected, which leads to a non-trivial coupling between the fields. Therefore, we do not try to analyze this correlator precisely in the UV-completed setup of this section.

However, we can make interesting remarks. Namely, in the limit $\omega \to 0$, taking $\xi = \zeta = \mathrm{const.}$ gives an exact solution to~\eqref{eq:xieq} and~\eqref{eq:zetaeq}\footnote{Note that strictly speaking, the decomposition of the gravity fluctuation modes in Appendix~\ref{app:fluctuations} only makes sense at non-zero mass of the fluctuations. The case of zero mass is special~\cite{Kiritsis:2006ua}. Therefore this massless solution only makes sense as the limiting solution $\omega \to 0$ of the massive solution.}. While we do not specify the action in the IR region, we expect that reasonable choices lead to this being the precise IR-normalizable solution. This means that the solution only has source terms in the limit $\omega \to 0$, so that the correlator must vanish in this limit. Further, the IR boundary condition in our language (e.g., that given in~\eqref{eq:IRbcgenqq}), which picks this solution is the Neumann condition, $\mathcal{A}=0$. That is, apart from leading to light modes, the Neumann boundary condition is singled out by this particular constant solution. Since the Neumann boundary conditions in general seem to lead to light modes, this gives further support to the presence of such modes in a properly IR completed setup.

\subsection{Flavor effects in Green's functions and identities}

Next we discuss various one-point functions and their identities.
Let us first analyze the free energy. Its variation satisfies 
\be \label{eq:dFexp}
 \delta F = -\frac{2\mathcal{N}\ell^3}{\kappa^2} \sigma \delta m_q
\ee
and in principle a similar term from the gluon sector proportional to the variation of the source $\phi_s$ in~\eqref{eq:VphiUVtext}. However we want to consider only variations that keep the source, and hence $r_\uv$, fixed. In this case the contribution from the gluon sector vanishes.
Inserting the expressions~\eqref{eq:mqspiralgen} and~\eqref{eq:sigmaspiralgen} in the differential
and integrating, we find a similar result as in the toy model,
\be \label{eq:Fonspiral}
\frac{\kappa^2}{2\mathcal{N}\ell^3}\left( F-F_0\right) = -\frac{1}{2} m_q \sigma  -\frac{X_\ir^2 \nu }{8 C_1^{(X)} C_2^{(X)} \sin\left(\rho_1-\rho_2\right) r_\ir^4 } =  -\frac{1}{2} m_q \sigma -\frac{X_\ir^2\ell_*^3 \nu^2 }{16 \ell^3 r_\ir^4 } \ ,
\ee

where  we used the identity in~\eqref{eq:Widentity} to obtain the latter expression. That is, this identity shows that the latter term is negative as was the case for the toy model in~\eqref{eq:freeencomp}. Therefore at zero quark mass, the solution with the smallest amount of scale separation (i.e., largest $r_\ir$) has always the lowest free energy, and as in the case of the toy model the same is typically found also for non-zero mass.
In~\eqref{eq:Fonspiral} we also included the constant of integration $F_0$, which is interpreted as the free energy of the solution in the origin of the spiral. This solution describes a flow from the UV fixed point to the IR fixed point, which is unperturbed by the flavors. Moreover, it may appear surprising that the result for the free energy depends on the IR scale $r_\ir$, which was defined in a somewhat arbitrary way in~\eqref{eq:rIRdef}. Note however that the dependence is only through the combination $X_\ir/r_\ir^2$, which is actually the combination of~\eqref{eq:IRinvar} where the dependence on the definitions of $r_\ir$ cancels.

In addition to integrating~\eqref{eq:dFexp}, the free energy and the energy momentum tensor can be computed either by evaluating the renormalized on-shell action or varying it with respect to the sources for the metric. We carry out both in Appendix~\ref{app:holo_renorm}, and find that the free energy density is given as
\be \label{eq:Fbdry}
 F = \frac{1}{4} \langle\theta^\mu_\mu\rangle = -\frac{\ell^3}{2\kappa^2}\left[\mathcal{N}m_q\sigma+(4-\Delta_g)\Delta_g\phi_s\phi_v\right] \ ,
\ee
which is consistent with $\langle\theta^{\mu\nu}\rangle = \eta^{\mu\nu}F$. Here the source and the vev for the dilaton were defined in~\eqref{eq:phisphiv}.
We already computed the free energy density on the Efimov spiral in~\eqref{eq:Fonspiral}. Combining with the exact expression~\eqref{eq:Fbdry} we can solve the gluon condensate contribution on the spiral: 
\be
 \frac{\ell^3}{2\kappa^2}(4-\Delta_g)\Delta_g\phi_s\phi_v = \frac{\mathcal{N}\ell^3}{2\kappa^2}m_q \sigma + \frac{\mathcal{N}\ell_*^3}{8\kappa^2}\frac{X_\ir^2\nu^2}{r_\ir^4} -F_0 \ ,
\ee
with $m_q$ and $\sigma$ given explicitly in~\eqref{eq:mqspiralgen} and~\eqref{eq:sigmaspiralgen}, respectively. This result captures the oscillation of the gluon condensate due to the backreaction of the flavor.
We stress that this expression only holds assuming a large amount of scaling, $r_\uv \ll r_\ir$, but no other assumptions are necessary.

The chiral condensate can be defined by
\be 
 \langle \bar q q\rangle = \frac{\partial F}{\partial m_q} = -\frac{2}{\kappa^2}\mathcal{N}\ell^3\sigma \ ,
\ee
where the summation over the flavor indices of the quark fields $\bar{q}$ and $q$ is assumed, and the last expression could be read off from~\eqref{eq:dFexp} or computed using the dictionary following the steps in Appendix~\ref{app:holo_renorm}. In Appendix~\ref{app:RG_flow} we also analyze the latter term of~\eqref{eq:Fbdry}, finding that
\be
 -\frac{\ell^3}{2\kappa^2}(4-\Delta_g)\Delta_g\phi_s\phi_v = \frac{N_c}{8\lambda^2} \beta(\lambda) \langle\mathrm{Tr}\,G_{\mu\nu}G^{\mu\nu}\rangle \ ,
\ee
where $\lambda = e^{\phi}$ is identified as the 't Hooft coupling as $r \to 0$, and $\beta(\lambda)$ is the $\beta$-function for this coupling.
Combining the results gives the Ward identity
\be \label{eq:Ward}
 \langle\theta^\mu_\mu\rangle =  m_q \langle \bar q q\rangle + \frac{N_c}{2\lambda^2} \beta(\lambda) \langle\mathrm{Tr}\,G_{\mu\nu}G^{\mu\nu}\rangle \ .
\ee

Note that the analysis presented here assumed $\Delta_g<4$, while the free field theory result suggests that we should set $\Delta_g=4$ at the boundary. This case cannot be obtained by simply setting $\Delta_g=4$ in the formulas above: holographic flows ending at $\Delta_g =4 $ typically involve logarithmic flows of the 't Hooft coupling $\lambda$. We analyze such flows in Appendix~\ref{app:RG_flow} and show that the Ward identity~\eqref{eq:Ward} continues to hold.

Finally, some additional information about the link between the near-boundary and walking solutions can be obtained by using the Wroskian relation~\eqref{eq:Widentity}. To see this, we write the full solution for $X$ as
\be
 X(r) = m_q r_\mathrm{UV} X_1(r) + \sigma r_\mathrm{UV}^3 X_2(r) = C_{\mathrm{tot}}^{(X)} \left(\frac{r}{r_\mathrm{UV}}\right)^2 \sin\left(\nu\log\frac{r}{r_\mathrm{UV}}+\rho^{(\mathrm{tot})}\right)
\ee
where the constants $C_{\mathrm{tot}}^{(X)}$ and $\rho^{(\mathrm{tot})}$ are defined by this equation and depend on $m_q$ and $\sigma$.
Then a direct calculation, using~\eqref{eq:Xwalkingcomps} gives for the variation of the phase as we move in the space of regular solutions (e.g. varying $m_q$)
\begin{align}
 \left(C_{\mathrm{tot}}^{(X)}\right)^2 \delta\rho^{(\mathrm{tot})} &= C_1^{(X)} C_2^{(X)} \sin\left(\rho_1-\rho_2\right) \left(\sigma \delta m_q - m_q \delta\sigma\right) & \\
 & = \frac{2\ell^3}{\nu \ell_*^3} \sigma^2 \delta\left( \frac{m_q}{\sigma}\right)
\end{align}
where we used the identity~\eqref{eq:Widentity} to obtain the expression on the second line.
If the quark mass is small, $|m_q| r_\mathrm{UV} \ll |\sigma| r_\mathrm{UV}^3$, this relation simplifies to
\be
  \left(C_{\mathrm{tot}}^{(X)}\right)^2 \delta\rho^{(\mathrm{tot})} = \frac{2\ell^3}{\nu \ell_*^3} \sigma \delta m_q \ .
\ee
We therefore find a simple expression for the variation of the free energy in terms of the walking region parameters:
\be
 \delta F = - \frac{\mathcal{N}\ell_*^3\nu}{\kappa^2}\left(C_{\mathrm{tot}}^{(X)}\right)^2 \delta\rho^{(\mathrm{tot})} 
\ee  
in the limit of small quark mass.

\section{Conclusion}
 
We have studied the dynamics of gauge theories near the conformal edge in the context of gauge/gravity duality. In particular, we have analyzed in detail the properties of the PNGB dilaton, which is generated when the scale symmetry of the near-conformal gauge theory is spontaneously broken by condensations of a quark-bilinear and gluons as well. For our analysis, we considered a gravity dual based on a five-dimensional geometry, which is asymptotically Anti-de Sitter in the deep UV. We also included bi-fundamental bulk scalars and a dilaton field, dual to the quark bilinears, and the trace of the gluon field strength squared, respectively. 
\\

When the gauge theory is near conformal, the anomalous dimension of the quark bilinear is close to one. This is reflected in the analysis of the gravity dual such that the bulk scalar violates the BF bound slightly. The IR geometry of the gravity dual for confining gauge theories, in general, deviates a lot from the AdS space, capping off in the IR for example. We cut off the space at the IR boundary where the walking behavior stops, while introducing a general, but linear, boundary condition at the IR boundary for the bulk scalars. We found that the PNGB dilaton is parametrically light compared to all other hadrons 
when the IR boundary condition is almost Neumann. The PNGB dilaton is a diffeomorphism-invariant mixture of a bulk scalar and the dilaton (or the scalar graviton). However, we find that the light mode is dominantly driven by the fluctuations of the scalar, suggesting that it is mostly a mesonic (rather than a glueball) state. This is in agreement with previous  studies in related models~\cite{Kutasov:2011fr,Arean:2013tja,Pomarol:2019aae}. We have checked for consistency that the PNGB dilaton saturates the anomalous Ward identity of the scale symmetry.  We also found that the hypothesis of the partially conserved dilatation current (PCDC) imposes a rather strong constraint on IR parameters to be consistent with the scale anomaly calculation in field theory by Gusynin and Miransky. We also calculated, for the case of a small quark mass, the pion mass and its decay constant in the gravity dual to find that the Gell-Mann-Oakes-Renner relation holds.

Our analysis was based on a rather generic bottom-up holographic model. Importantly, unlike many previous studies in the literature, apart from the physics near the IR fixed point implementing the conformal behavior, we also considered the effects of the full RG flow. That is, we considered the flow of the coupling from the IR fixed point to the UV fixed point (modeling free theory) at high energies (UV), and the physics of breaking conformal and chiral symmetries at low energies (IR). As for the UV flow, our analysis is complete in the sense that it captures all relevant flows in the class of models we are considering. For the UV behavior, our general observations were further supported by a simple analytic ``toy model'' of the UV physics in Sec.~\ref{sec:toymodel}. 

However, IR physics is, in general, more complicated. In the holographic model, the scalar fields grow large in the IR region, meaning that non-linearities and backreaction effects become important. This reflects the complexity of the IR physics in field theory. In this work, we implemented the IR physics by considering generic IR boundary conditions for the fields and their fluctuations. We argued that such boundary conditions are indeed capable of capturing generic IR physics. We observed, in agreement with earlier literature (see, e.g.,~\cite{Arean:2013tja,Elander:2018gte}), that while large scale separation leads to special scalar modes, whether they are (anomalously) light in the end depends on the IR boundary conditions. 

It would be important to check that our findings hold in a model that (apart from the UV flow) explicitly implements an IR completion following some of the examples in the literature, for example~\cite{Jarvinen:2011qe,Kutasov:2012uq,Pomarol:2019aae}. Such an IR completion should be general enough to capture all possibilities considered here, in particular, the cases which show large scale separations between the UV and IR that differ from the standard ``Miransky'' scaling law. Note that it was in such cases with non-standard scaling, and with Neumann boundary conditions for the fluctuations, where we found an anomalously light scalar mode in our analysis. Explicit modeling of the IR physics should confirm that it is indeed possible to obtain this kind of mixture (both for the background and the fluctuations). We already presented an argument based on the analysis of the fluctuation equations in Sec.~\ref{sec:flucts} that the Neumann boundary conditions will be singled out by a full model, but this should also be confirmed by explicit solutions. Having a full model of the IR dynamics would also allow us to study the mixing of the meson and glueball modes in the IR region explicitly.

\acknowledgments

This work was supported by the National Research Foundation of Korea (NRF) grants
funded by the Korean government (MSIT) (grant numbers 2021R1A4A5031460 (DKH)
and 2021R1A2C1010834 (MJ)), and was also supported by IBS under the project code,
IBS-R018-D1 (SHI). DKH acknowledges the support from Basic Science Research Program through the National Research Foundation of Korea (NRF) funded by the Ministry of Education (NRF-2017R1D1A1B06033701). JCR and MJ have been supported by an appointment to the JRG
Program at the APCTP through the Science and Technology Promotion Fund and Lottery
Fund of the Korean Government. JCR and MJ have also been supported by the Korean
Local Governments -- Gyeongsangbuk-do Province and Pohang City. JCR was additionally supported by a DGAPA-UNAM
postdoctoral fellowship.


\newpage 
\appendix

\setcounter{footnote}{0}
\renewcommand{\thefootnote}{\arabic{footnote}}

\section{Precise dictionary between gravity and field theory}

In this Appendix, we review the dictionary between the holographic model and the gauge theory. 
We first construct the relevant boundary term on the QCD side of the gauge/gravity correspondence, which fixes how the source terms of gravity couple to the operators of field theory.
We start by considering the QCD action generalized to a background with the metric $\gamma$ which is assumed to be close to the Minkowski metric:
\be
 S_\gamma = \int d^4x\,\sqrt{-\det\gamma}\bigg [
-{1\over2g^2}\,{\mathrm{Tr}} G_{\mu\nu}G_{\r\s}\gamma^{\m\r}\gamma^{\n\s}+i \gamma^{\m\n}\bar q \gamma_\m D_\n q
-\bar q_R\,M_q\,q_L-\bar q_L\,M_q^\dagger\,q_R
\bigg]\,,
\ee
where $q_L = (1 +\gamma^5)\,q/2$, $q_R = (1 -\gamma^5)\,q/2$, and $M_q$ is the (potentially complex) quark mass matrix.

The energy-momentum tensor (assuming mostly minus conventions) is defined as
\be
 \theta^{\m\n} = - 2 \left.\frac{\delta S_\gamma}{\delta \gamma_{\m\n}}\right|_{\gamma=\eta}
\ee
where $\eta$ denotes the Minkowski metric $\eta =\mathrm{diag}(1,-1,-1,-1)$. Therefore varying around the Minkowski metric, i.e., taking $\gamma_{\m\n} = \eta_{\m\n} + \delta \gamma_{\m\n} + \cdots$, we find that 
\be
 S_\gamma = S_\mathrm{QCD} - \frac{1}{2} \int d^4 x\ \theta^{\mu\nu}\ \delta \gamma_{\m\n} + \cdots
\ee
where the dots denote higher order variations.

The precise identification of the source term for the bulk 5D gravity is most conveniently done (assuming that the background metric will be asymptotically AdS$_5$ with power law corrections) in Fefferman-Graham coordinates. The Poincar\'e covariant background metric can be written as
\be \label{eq:FGmetric}
 ds^2 = \frac{\ell^2}{z^2}(dz^2 - e^{2 \bar A(z)} \eta_{\m\n}dx^\m dx^\n)
\ee
where $\ell$ is the AdS radius and the warp factor $\bar{A}$ depends on the gravity theory. AdS$_5$ asymptotics requires $\bar{A} \to 0$ as $z \to 0$. 

Allowing for a more general fluctuation solution, which also depens on the space-time coordinates, the standard near boundary expansion is 
\bea
 ds^2 &=& \frac{\ell^2}{z^2}\left[dz^2 - \left(g^{(0)}(x)_{\m\n} + g^{(2)}(x)_{\m\n} z^2 +g^{(4)}(x)_{\m\n} z^4 + \cdots\right) dx^\m dx^\n\right] \\
 &\equiv & \frac{\ell^2}{z^2}\left[dz^2 - \hat g_{\m\n} dx^\m dx^\n\right] \ . 
\eea
The link to field theory is then obtained by identifying 
\be
 g^{(0)}_{\m\n} =\gamma_{\m\n} = \eta_{\m\n} + \delta \gamma_{\m\n} + \cdots\ .
\ee

We also need the dictionary for the other operators. These are the dilaton field $\phi$ dual to the ${\mathrm{Tr}}\,G_{\mu\nu}^2$, and the complex scalar $X$ which is a matrix in flavor space and dual to $\bar q q$. The simplest possibility is that for asymptotically AdS$_5$ spaces the dilaton approaches a constant at the boundary, which is identified as the source for the  $\mathrm{Tr}\,G_{\mu\nu}^2$ operator, i.e., the gauge coupling. If the dimension of the operator $\Delta_g$ is less than four, we can write $\phi(z,x) = \phi_0(x,z) + \morder{z^{2\Delta_g-4}} = \phi_s(x) z^{4-\Delta_g} + \morder{z^2} $. We consider more complicated cases below in Appendix~\ref{app:RG_flow}. The flavor field $X$ is assumed to have the asymptotics near the boundary
\be
 X(z,x) = X_-(x)\ z^{4-\Delta}(1+\morder{z^2}) + X_+(x)\ z^{\Delta}(1+\morder{z^2})
\ee
with $3\ge \Delta >2$. 

After these identifications we are ready to write the complete boundary term at $z=\epsilon\ll1$. It is given by
\bea \label{eq:Sbdry}
 S_\mathrm{bdry}\!\! &=&\!\! \int d^4 x\bigg[- \frac{1}{2} \theta^{\mu\nu} \delta \hat g_{\m\n} -\frac{N_c}{2} e^{-\phi}\ \mathrm{Tr} G_{\mu\nu}G^{\mu\nu}
 -\frac{K_X}{z^{\Delta-4}}\bar q_R\,X \,q_L-\frac{\bar K_X}{z^{\Delta-4}}\bar q_L\,X^\dagger\,q_R\bigg]_{z=\eps} \nonumber\\
 \!\!&=&\!\!\int d^4 x\bigg[- \frac{1}{2} \theta^{\mu\nu} \delta \gamma_{\m\n} -\frac{N_c}{2} e^{-\phi_0}\ \mathrm{Tr} G_{\mu\nu}G^{\mu\nu}
 -K_X\bar q_R\,X_- \,q_L-\bar K_X\bar q_L\,X_-^\dagger\,q_R\bigg]\nonumber\\&&\times \left[1+\morder{\epsilon^2}+\morder{\epsilon^{2\Delta-4}}\right] 
\eea
The interpretation is that the UV cutoff $\eps$ maps to the (inverse of the) renormalization scale of field theory. Here $\delta \hat g_{\m\n} = \hat g_{\m\n} -\eta_{\m\n}$ and $K_X$ is a constant whose value cannot be determined. This reflects the fact that the quark mass is not an observable. In the analysis in this article we will set the constant to be one for simplicity. 
Note that the latter three terms also formally contain the kinetic term of the gauge field as well as the fermion mass terms in the QCD action. In particular, we identify $X_-$ as the quark mass matrix, $M_q = X_-$. Below we will only consider the case where both $X_-$ and $M_q$ are proportional to the identity matrix in flavor space. In this case, we denote the (real) quark mass as $m_q$.

Let us then comment on the five-dimensional gravity action. It is given by various terms as follows:
\be \label{eq:S5dfull}
 S = S_\mathrm{grav} + S_\mathrm{matter} + S_\mathrm{GH} + S_\mathrm{ct}
\ee
where the first two terms are the five-dimensional terms given in the text for various models (e.g.~\eqref{eq:gravSc},~\eqref{eq:Xaction}). The third term is the Gibbons-Hawking boundary term
\be
 S_\mathrm{GH} = \frac{1}{\kappa^2} \int d^4x \sqrt{-\det \gamma_b} K
\ee
where $K$ is the extrinsic curvature of the boundary metric $\gamma_b$, which is the pullback of the full metric $g$ (including the factors $\ell/z$) on the surface $z=\epsilon$. This term is required in order for the variational problem and thermodynamics of the setup to be consistent. 
The last term in~\eqref{eq:S5dfull} is the four dimensional boundary counterterm which will be given below, and cancels the UV divergences in the gravity and matter terms.

Finally, the dictionary between the field theory and gravity is obtained from
\be
 Z_\mathrm{QCD} \equiv \exp\left[iS_\mathrm{QCD}+iS_\mathrm{bdry}\right] = Z_\mathrm{gr} \equiv \exp\left[iS\right]
\ee
with the understanding that the sources appearing in~\eqref{eq:Sbdry} are set to be the same on both sides of the duality.

\section{Invariant fluctuations and fluctuation equations}\label{app:fluctuations}

In order to study the correlators, we need to write down the definitions for the various fluctuations. For the metric we use
\be
 g_{zz} = \frac{\ell^2}{z^2}(1+2\hat\phi(z,x)) \ , \quad g_{z\mu} = \frac{\ell^2}{z^2}e^{\bar A(z)}\hat A_\mu(z,x)\ , \quad g_{\m\n} = \frac{\ell^2}{z^2}e^{2\bar A(z)}(\eta_{\m\n}+h_{\m\n}(z,x))
\ee
and for the scalars
\be
 \phi(z,x) = \bar \phi(z) + \varphi(z,x) \ , \qquad X_{ij}(z,x) = \delta_{ij}\bar X(z) + \delta_{ij}\chi(z,x) \ .
\ee
Here $i,j$ are the flavor indices. We only consider the flavor independent case so we can take the flavor background and fluctuations to be proportional to the unit matrix. We focus here on the scalar (spin zero) sector. This sector has two dynamical fields. It is convenient to use the diffeomorphism invariant combinations. To do this we define
\be
 \psi = \frac{1}{6}\left(h^\m_\m - \frac{\partial^\m \partial^\n }{\partial^2}h_{\m\n}\right) \ , \qquad \tau = \frac{1}{2}\frac{\partial^\m \partial^\n }{\partial^2}h_{\m\n} \ , \qquad h^\m_\m = 2(3\psi+\tau) \label{psidef}
\ee
Conversely, the metric fluctuations are written in terms of these fields as
\be \label{eq:hdecomp}
 h_{\m\n} = 2\left(\eta_{\m\n}- \frac{\partial_\m \partial_\n}{\partial^2} \right)\psi + 2 \frac{\partial_\m \partial_\n}{\partial^2} \tau + \sum_{i}P_{\m\n}^{(i)} s^{(i)}
\ee
where $s^{(i)}$ are the spin-one and spin-two degrees of freedom the projectors of which satisfy $P_{\m}^{\m\,(i)}=0 =\partial^\m \partial^\n P_{\m\n}^{(i)}$.
Notice that if we, for example, assume time-dependent homogenous fluctuations (which is enough to capture the massive modes), $\psi$ is proportional to the trace of the spatial components of the fluctuations.

It was simplest to write the boundary dictionary in the Fefferman-Graham coordinates defined in~\eqref{eq:FGmetric}.
As it turns out, however, for the full picture it is more convenient to use the conformal coordinates than the Fefferman-Graham coordinates. This is because the Fefferman-Graham conventions explicitly fix the value of the AdS radius, and the full flow will have different AdS radii at the UV and IR fixed points, so the FG conventions can only be convenient near one of the fixed points. The change of coordinates is defined by
\be
 ds^2 = \frac{\ell^2}{z^2}(dz^2 - e^{2 \bar A(z)} \eta_{\m\n}dx^\m dx^\n) = e^{2 A(r)}(dr^2 -\eta_{\mu\nu}dx^\mu dx^\nu)
\ee
so that $A = \bar A-\log z +\log \ell$ and 
\be
 \frac{dr}{dz} = \frac{\ell}{e^{A}z} = e^{-\bar A}\ .
\ee
We will be using the conformal coordinates in the rest of this Appendix and elsewhere in the article.

The gauge invariant (i.e., diffeomorphism invariant) combinations are written in conformal coordinates as
\be
 \zeta = \psi - \frac{A'}{\bar \phi'} \varphi \ , \qquad \xi = \psi - \frac{A'}{\bar X'} \chi \label{giv}
\ee

The first-order fluctuation action around the background solution is always a boundary term. It may be written as
\be
 S_\mathrm{ren}^{(1)} = S_\mathrm{bdry}^{(1)}+S_\mathrm{ct}^{(1)} \ .
\ee
In the ``standard gauge'' where $\hat \phi=0$ and $\partial^\m \hat A_\m=0$, the first term simplifies to
\be \label{eq:S1nonre}
 S^{(1)}_\mathrm{bdry} = \frac{e^{3 A(r)}}{\kappa^2\ell}\int d^4x\,\left[ 3\ell A'(r)\left(3\psi(r,x)+\tau(r,x)\right)+\mathcal{N} \ell X'(r) \chi(r,x)\right]\Big|_{r=\eps}
\ee
where $\eps$ is the value of the UV cutoff. This term is UV divergent and needs to be renormalized by adding the correct boundary counterterms contribution $S_\mathrm{ct}^{(1)}$ as we discuss in Appendix~\ref{app:holo_renorm}.

The renormalized second-order action is written as 
\be \label{eq:S2ren}
 S^{(2)}_\mathrm{ren} = S^{(2)}_\mathrm{bdry}+S^{(2)}_\mathrm{ct} + S^{(2)}_\mathrm{bulk}
\ee
where the first term is a finite boundary term
(see, e.g.,~\cite{Kofman:2004tk}), the second term is the counterterm which will be specified below, and 
\bea
 S^{(2)}_\mathrm{bulk} &=& - \frac{1}{2\kappa^2}\int d^5x\sqrt{-\det \bar g}\  \bigg[ \frac{(\phi')^2}{(A')^2}\ \bar g^{MN}\partial_M\zeta\partial_N \zeta + \mathcal{N} \frac{(\bar X')^2}{(A')^2}\ \bar g^{MN}\partial_M\xi\partial_N \xi\nonumber \\
  &&- \frac{\mathcal{N}e^{2A} }{A'^2} \mathcal{M}\, (\xi-\zeta)^2 \bigg]  
\eea
where $\bar g_{MN}$ is the background metric, and
\begin{align}
\label{eq:Mrdef}\nonumber
{\mathcal{M}}(r) &= -e^{2A(r)}X'(r) \phi '(r)\bigg[2 X(r) m_X(\phi (r)) m_X'(\phi (r)) \left(1+\frac{\mathcal{N} X(r) X'(r)}{3 A'(r)}\right)
 &\\ \nonumber
&+2 X(r) m_X(\phi (r))^2  \left(1+\frac{\mathcal{N} X(r) X'(r)}{3 A'(r)}\right)\frac{\phi'(r)}{3 A'(r)}+\frac{X'(r) V'(\phi (r))}{3  A'(r)} &\nonumber\\
&+\frac{2 V(\phi (r)) X'(r) \phi '(r)}{9  A'(r)^2}\bigg]&
\end{align}

The fluctuation equations following from this action read 
\begin{align}
\label{eq:xieqA}
 \xi''(r) + \partial_r \log \frac{e^{3 A(r)} X'(r)^2}{A'(r)^2}\ \xi'(r) 
 +\frac{e^{2A(r)}}{X'(r)^2}{\mathcal{M}}(r)(\xi(r)-\zeta(r)) +\omega^2\xi(r)
 &= 0& \\
\label{eq:zetaeqA}
 \zeta''(r) + \partial_r \log \frac{e^{3 A(r)} \phi'(r)^2}{ A'(r)^2}\ \zeta'(r) 
  +\frac{\mathcal{N}e^{2 A(r)}}{\phi'(r)^2}{\mathcal{M}}(r)(\zeta(r)-\xi(r)) +\omega^2\zeta(r)
 &= 0&  
\end{align}

For the purpose of this article, we can restrict to the region where the background field $X$ is small. More precisely (following definitions of the main text) whenever $r \ll r_\ir$, the bulk mass term in the fluctuation equation for $\zeta$ becomes subleading (i.e., it is $\sim X^2 \sim (r_\uv/r_\ir)^2$) and can be dropped, so that
\begin{align}
\label{eq:xieq2A}
 \xi''(r) + \partial_r \log \frac{e^{3 A(r)} X'(r)^2}{ A'(r)^2}\ \xi'(r)  
  +\frac{e^{2A(r)}}{X'(r)^2}{\mathcal{M}}(r)(\xi(r)-\zeta(r))
+\omega^2\xi(r)
 &= 0& \\
\label{eq:zetaeq2A}
 \zeta''(r) + \partial_r \log \frac{e^{3 A(r)} \phi'(r)^2}{A'(r)^2}\ \zeta'(r) 
+\omega^2\zeta(r)
 &= 0 &  
\end{align}
That is, the smallness of the field $X$ dynamically ensures that we are working in the probe limit, where the gravity fluctuation $\zeta$ and the matter fluctuation $\xi$ partially decouple. This means that as for computing te spectrum, the equations for $\zeta$ and $\xi$ (after setting $\xi=0$ there) can be analyzed separately. However the remaining mixing term in the equations can be relevant for expressions for the correlators.

\section{Equations of Motion} \label{app:eoms}

In this appendix we present for completeness the full equations of motion arising from the action \eqref{fullaction}.

We first write down the equations of motion for the dilaton $\phi$ and the field  $X$.
\begin{align}\label{dilatoneom}
\nabla^2 \phi + \frac{1}{2}\frac{\partial}{\partial \phi}\Big(V(\phi)+\mathcal{N}m_X(\phi)^2 X^2 \Big)=0
\end{align}
\begin{align}\label{Xeom}
\nabla^2 X +\mathcal{N}m_X(\phi)^2 X =0
\end{align}
where $\nabla$ is the covariant derivative with respect to \eqref{eq:metric}. The Einstein equations take the form
\begin{equation}
R_{M N}-\frac{1}{2} g_{M N} R=T_{M N}^g+T_{M N}^f
\end{equation}
where
\begin{align}
&T_{M N}^g=\nabla_M \phi \nabla_N \phi-\frac{1}{2}g_{MN}\nabla_P\phi \nabla^P\phi +\frac{1}{2}V(\phi)& \\
&T_{M N}^f=\mathcal{N}\Big(\nabla_M X \nabla_N X-\frac{1}{2}g_{MN}\nabla_P X \nabla^P X \frac{1}{2}g_{MN}m_X(\phi)^2 X^2\Big)
\end{align}

We then write down the equations of motion for the background, i.e., for the metric in~\eqref{eq:metric}. The equations can be written as
\begin{align}
-e^{2A}\big( V(\phi)+ \mathcal{N}m_X(\phi)^2 X^2\big)+12A'^2-\mathcal{N}X'^2-\phi'^2=0
\end{align}
\begin{align}
-e^{2A}\big( V(\phi)+ \mathcal{N}m_X(\phi)^2 X^2\big)+\mathcal{N}X'^2+\phi'^2+6A'^2+6A''=0
\end{align}
\begin{align}\label{dilatoneombg}
e^{-2A}\big(\phi''+3A'\phi' \big)+\mathcal{N}m_X(\phi)^2 m_X'(\phi) X^2+\frac{1}{2}V'(\phi)=0
\end{align}
\begin{align}\label{Xeombg}
e^{-2A}\big(X''+3A'X' \big)+m_X(\phi)^2 X=0
\end{align}

\section{Expansions near the UV boundary} \label{app:uvasympt}

In order to analyze the UV divergence of the action, we need the expansions of the background and the fluctuations near the boundary. We start from the background.
By using the equations of motion from Appendix~\ref{app:eoms}, it is found that they obey the following asymptotics at the boundary:
\bea
 \bar X(r) &\sim& X_-\ r^{4-\Delta} + X_+\ r^{\Delta} \\
 \label{eq:AUVexp}
 A(r) &\sim& \log \frac{\ell}{r}-\frac{\mathcal{N}(\Delta-4)X_+^2}{6(2\Delta-9)} r^{8-2 \Delta }-\frac{\mathcal{N} X_+ X_- }{30} \Delta  (4-\Delta) r^4\\
&& -\frac{\mathcal{N} X_+^2 \Delta}{6}  r^{2 \Delta } \nonumber
\eea
where $\Delta=\Delta_\uv$ is the UV dimension for the field $X$ defined in~\eqref{eq:FPmasses}. 
Each term has power law suppressed corrections that were not written down. Notice that there is no VEV term (apart from that caused by direct backreaction of the flavors) for the metric:  this is not an assumption but follows from the Einstein equations. 

Without loss of generality we can assume that the value of the dilaton at the boundary vanishes, $\phi_\mathrm{UV}=0$. Further, assuming a simple mass term for the dilaton, i.e.,
\be
\label{eq:VphiUV}
 V(\phi) = \frac{12}{\ell^2} + \frac{\Delta_g(4-\Delta_g)}{\ell^2}\phi^2 + \morder{\phi^4}
\ee
the boundary expansion for the dilaton takes the standard form,
\be
 \phi(r) = \phi_s r^{4-\Delta_g} + \phi_v r^{\Delta_g} + \cdots
\ee
Near the boundary the behavior of the dilaton and the flavor field $X$ are the same up to the change in the dimension. Because of this, the contribution from the dilaton and $X$ to most of the correlators take the same form in most of the results below. Consequently, we often omit the contributions from the dilaton for brevity. 
We will be most interested in the contributions to the Green's functions from the flavor sector, and the contributions from the dilaton are analogous to those. In particular, the field $\zeta$ has analogous expansion as the field $\xi$. If we however set $\Delta_g=4$ at the boundary, so that the mass term in~\eqref{eq:VphiUV} vanishes, the expansion is more complicated (see Appendix~\ref{app:RG_flow}). The source term is replaced by a logarithmic flow. Asymptotics in this class have been considered in the literature e.g. in the context of the improved holographic QCD model~\cite{Gursoy:2007cb,Gursoy:2007er,Jarvinen:2015ofa}, but as we are mostly interested in the flavor terms here, we will simply set the dilaton to zero for the UV expansions.

As it turns out, it is not possible to write expansions for all fluctuations in powers of $z$ that would be well behaved for all $2<\Delta<3$. 
Restricting to fluctuations only depending on the holographic coordinate (spatial and time dependence will only affect higher order terms), all relevant terms in the UV are however included in the following integral expressions:
\bea
\xi &=& \xi_- + \tilde \xi_+ \int_0^r d\tilde r \frac{\ell ^3 e^{-3A} A'(\tilde r)^2}{X'(\tilde r)^2} \\ \nonumber
&\sim &\xi_- +\xi_+\left[\frac{r^{2\Delta-4}}{1+\frac{\Delta  X_+ r^{2 \Delta -4}}{(4-\Delta) X_-}} - \frac{ (\Delta -4)(\Delta -2) (4 \Delta -19) \mathcal{N} X_-^2}{ (6\Delta-27 )}\, F_1(r^{2\Delta-4})\, r^4\right] 
\eea
where $\xi_+ = \tilde \xi_+/((4-\Delta)^2(2\Delta-4)X_-^2)$ and 
\bea
 F_1(r^{2\Delta-4}) &=&4 \int_0^1 \frac{dw}{w} \left(1+\frac{\Delta X_+r^{2 \Delta -4}w^{2 \Delta -4}}{(4-\Delta) X_-}\right)^{-2}\\ \nonumber
&&\times \left(w^4+\frac{11(2\Delta -9) \Delta  X_+ w^{2 \Delta }}{(4\Delta-19)5 X_-}r^{2 \Delta -4}-\frac{\Delta(4 \Delta+3) (2 \Delta -9) X_+^2 w^{4 \Delta-4 }}{(\Delta-4)(2\Delta+1)(4\Delta-19)  X_-^2 }r^{4\Delta-8}\right)
\eea
In the standard gauge (i.e.,  $\hat \phi=0$ and $\partial^\m \hat A_\m=0$), we find for the gauge non-invariant functions
\bea
\chi(r) & = & \tilde \chi_- \ell e^{-A(r)} X'(r) - \tilde \xi_+  X'(r)\int_0^r d\tilde r\  \frac{\ell^3 A'(\tilde r))}{e^{3A(\tilde r)}X'(\tilde r)^2} \\ 
&\sim & \chi_- r^{4-\Delta }\left(1+\frac{\Delta  X_+ r^{2 \Delta -4}}{(4-\Delta) X_-}\right)+ (4-\Delta)  \xi_+ X_- r^{\Delta } \\ 
\psi(r) & = & \xi_- +\ell e^{-A(r)}A'(r)\tilde \chi_- - \tilde \xi_+\int_0^r d\tilde r  \frac{\ell^3 A'(\tilde r)(e^{-A(r)}A'(r)-e^{-A(\tilde r)}A'(\tilde r))}{e^{2A(\tilde r)}X'(\tilde r)^2}  \\
&\sim & \psi_--\frac{1}{6} \mathcal{N} X_- \chi_- \left[ r^{8-2 \Delta }+\frac{\Delta  X_+ r^4}{X_-} +\frac{\Delta  X_+^2 r^{2 \Delta }}{(4-\Delta) X_-^2}\right] \nonumber\\
&&-\frac{(2-\Delta)(4-\Delta)^2\mathcal{N}X_-^2}{3}\xi_+ r^4 F_2(r^{2\Delta-4}) \\ 
\tau(r) & = & \tau_-+\psi(r)-\psi_- + \tilde \xi_+ \int_0^r d\tilde r \frac{\ell^3 ( A''(\tilde r)-A'(\tilde r)^2)}{e^{3A(\tilde r)}X'(\tilde r)^2} 
\eea
with
\be 
\psi_- = \xi_--\frac{\chi_-}{(4-\Delta) X_-} \ , \qquad \tilde \chi_- = \frac{\chi_-}{4-\Delta}
\ee
and
\bea
 F_2(r^{2\Delta-4}) &=& \int_0^1 \frac{dw}{w} \left(1+\frac{\Delta X_+r^{2 \Delta -4}w^{2 \Delta -4}}{(4-\Delta) X_-}\right)^{-2}\nonumber\\
&&\times \Bigg(\frac{\left(w^8-w^{2 \Delta }\right)}{(\Delta -4) w^4}+\frac{ \Delta  \left(w^4-1\right) X_+ w^{2 \Delta -4} r^{2 \Delta -4}}{(\Delta -4) X_-}\nonumber \\
&&\quad -\frac{ \Delta  X_+^2 \left(w^{2 \Delta }-1\right) w^{2 \Delta -4} r^{4 \Delta -8}}{(\Delta -4)^2 X_-^2} \Bigg)
\eea
The functions $F_{1,2}$ were normalized such that $F_1(0)=1=F_2(0)$. 
We omitted corrections  suppressed by integer powers of $z$ to each term which also contain non-trivial dependence on space-time coordinates. 
Notice that there are three independent sources ($\chi_-$, $\psi_-$, $\tau_-$) but only one vev coefficient $\xi_+$.

\section{Holographic renormalization}\label{app:holo_renorm}

We will carry out holographic renormalization by using the counterterm method. That is, we write the covariant boundary terms which cancel the UV divergences in the on-shell gravity action. Here we will assume, for simplicity $2<\Delta<3$. For $3<\Delta<4$, additional counterterms would be needed.

The counterterms are given by \cite{Papadimitriou:2004ap, Papadimitriou:2011qb}
\bea
 S_\mathrm{ct} &=& - \frac{1}{2\kappa^2} \frac{\ell}{2}\int d^4x\,\sqrt{-\det \gamma_b}\left(\frac{12}{\ell^2}+R_\gamma\right)\bigg|_{r=\eps} \nonumber\\
 &&-\frac{(4-\Delta)\mathcal{N}}{2\kappa^2\ell } \int d^4x\,\sqrt{-\det \gamma_b}\, X(r,x)^2\Big|_{r=\eps}
 \label{eq:Xct}
\eea
where $R_\gamma$ is the scalar curvature for the boundary metric $\gamma_b$. 
For the renormalization of the dilaton field, we need an additional counterterm
\be
 S_\mathrm{ct,\phi} = -\frac{(4-\Delta_g)}{2\kappa^2\ell } \int d^4x\,\sqrt{-\det \gamma_b}\, \phi(r,x)^2\Big|_{r=\eps} \ .
\ee
In principle, one should also add a derivative counterterm but this will not affect the analysis in this article.

After adding the counterterm contributions to~\eqref{eq:S1nonre}, the first-order fluctuation term becomes
\bea \label{eq:action1}
  S^{(1)}_\mathrm{ren} &=& \frac{1}{2\kappa^2}\frac{e^{4 A(r)}}{\ell}\int d^4x\,\Big[ - \left(\mathcal{N}(4-\Delta) X(r)^2+(4-\Delta_g) \phi(r)^2+6 \ell e^{- A(r)} A'(r)+6\right) \nonumber\\
&&  \times\left(3\psi(r,x)+\tau(r,x)\right) +2 \mathcal{N}  \left(\ell e^{- A(r)} X'(r)-(4-\Delta) X(r)\right)\chi (r,x)\\ 
&& +2  \left(\ell e^{- A(r)} \phi'(r)-(4-\Delta_g) \phi(r)\right)\varphi (r,x) \Big]\Big|_{r=\eps} \nonumber
\eea
which, as one can check by using the asymptotics from Appendix~\ref{app:uvasympt}, is divergence free.

After adding the second-order contribution in fluctuations $S^{(2)}_\mathrm{ct}$ from the counterterm~\eqref{eq:Xct}, the second-order action~\eqref{eq:S2ren} simplifies to
\bea
S_\mathrm{ren}^{(2)} &=& S_\mathrm{bulk}^{(2)} + \frac{1}{2\kappa^2}\frac{e^{4 A(r)}}{2\ell}\int d^4x\,\bigg[ \\&& \left(\mathcal{N}(4-\Delta) X(r)^2+6 \ell e^{-A(r)} A'(r)+6\right)
 \times \left(\tau (r,x)^2-6 \tau (r,x) \psi (r,x)-3 \psi (r,x)^2\right)\nonumber \\ 
 &&+4 \mathcal{N} \left(\ell e^{-A(r)} X'(r)+(\Delta -4) X(r)\right) \chi (r,x) (\tau (r,x)+3 \psi (r,x))\nonumber \\ 
 && + 2 \mathcal{N}  \left(\Delta-3 +\frac{\ell e^{-A(r)}}{X'(r)}\left(X''(r)-A'(r)X'(r)\right)\right)\chi (r,x)^2 
 \bigg]_{r=\eps}\nonumber
\eea
where we suppressed the dilaton contributions which are analogous to the contributions from $X$ and its fluctuation $\chi$.
In addition, we used the constraint in the standard gauge, 
\be
 3 \partial_r \psi(r,t)+ \mathcal{N} X'(r) \chi (r,t) = 0 
\ee
and carried out partial integrations in the space-time directions which were not made explicit.

\subsection{Correlators} \label{app:correlators} 

Let us restrict to the correlators of the trace $\theta_\mu^\mu$. In order to do this, we consider fluctuations with boundary conditions $\psi_- =\tau_-$, and $\chi_- = 0$ in the UV expansions given above, setting all other sources to zero. Then also $\xi_-=\psi_-$. 

Using~\eqref{eq:hdecomp} for these boundary conditions, the boundary term~\eqref{eq:Sbdry} becomes
\be
  S_\mathrm{bdry} =  -\int d^4x\ \theta^\mu_\mu \psi_-
\ee
Thus the generating functional is
\be
 Z_\mathrm{QCD}[\psi_-] = \int \mathcal{D} e^{iS_\mathrm{QCD} - i\int d^4x\ \theta^\mu_\mu \psi_- } 
\ee
The gauge/gravity correspondence is manifested as
\be
 \frac{Z_\mathrm{QCD}[\psi_-]}{Z_\mathrm{QCD}[0]} = \exp\left[ iS^{(1)}_\mathrm{ren}+iS^{(2)}_\mathrm{ren}\right]_\mathrm{on-shell}
\ee
with the understanding that the right hand side is evaluated for the regular solution with the above UV boundary conditions. 

Neglecting the dilaton contributions for a moment, for these terms we find, on-shell, that
\bea
 S^{(1)}_\mathrm{ren} &=& - \frac{2\ell^3}{\kappa^2\eps^4}\left(\mathcal{N}(4-\Delta) X(\eps)^2+6 \ell e^{-A(\eps)} A'(\eps)\right) \int d^4x\ \psi_-\nonumber \\ 
 &=&  \frac{2\ell^3}{\kappa^2} \mathcal{N}(\Delta-2)(4-\Delta)X_+X_- \int d^4x\ \psi_-\\
 S_\mathrm{bulk}^{(2)} &=& \frac{\mathcal{N}\ell^3}{2\kappa^2} X_-^2 (4-\Delta)^2(2\Delta-4)\int d^4x\ \psi_- \xi_+ \\
 S^{(2)}_\mathrm{bdry} +S^{(2)}_\mathrm{ct} &=& -\frac{2\ell^3}{\kappa^2\eps^4} \left(\mathcal{N}(4-\Delta) X(\eps)^2+6 \ell e^{-A(\eps)} A'(\eps)\right)\int d^4x\ \psi_-^2 \nonumber \\
 &=& \frac{2\ell^3}{\kappa^2} \mathcal{N}(\Delta-2)(4-\Delta)X_+X_-\int d^4x\ \psi_-^2 
\eea
where $S^{(2)}_\mathrm{ren} = S^{(2)}_\mathrm{bdry} +S^{(2)}_\mathrm{ct} + S_\mathrm{bulk}^{(2)}$.
Therefore we obtain for the correlators
\bea
 \langle\theta^\mu_\mu(x)\rangle &=& i \frac{\delta}{\delta \psi_-(x)}\log Z_\mathrm{QCD}[\psi_-]\Big|_{\psi_-=0} \nonumber\\
 &=& - \frac{2\ell^3}{\kappa^2} \mathcal{N}(\Delta-2)(4-\Delta)X_+X_- \\
 \langle\theta^\mu_\mu(x_1)\theta^\mu_\mu(x_2)\rangle &=& - \frac{\delta}{\delta \psi_-(x_1)}\frac{\delta}{\delta \psi_-(x_2)}\log Z_\mathrm{QCD}[\psi_-]\Big|_{\psi_-=0} \nonumber\\
 &=& - \frac{4\ell^3}{\kappa^2} \mathcal{N}(\Delta-2)(4-\Delta)X_+X_- \delta^{(4)}(x_1-x_2) \nonumber \\
 && - \frac{\mathcal{N}\ell^3}{2\kappa^2} X_-^2 (4-\Delta)^2(2\Delta-4) \left[\frac{\delta \xi_+(x_1)}{\delta \psi_-(x_2)}+\frac{\delta \xi_+(x_2)}{\delta \psi_-(x_1)}\right]
 \label{eq:twopointf}
\eea
where the last term needs to be calculated by solving the $\xi$ fluctuation equation with proper IR regularity condition. Since different Fourier modes decouple in the fluctuation equations, and the equations only depend on $q^2$ where $q_\mu$ is the momentum, the Fourier transformed coefficients $\delta \tilde \xi_+(q)$ and $\delta \tilde \psi_-(q)$ satisfy
\be
 \delta \tilde \xi_+(q) = \widetilde G(q) \delta \tilde \psi_-(q) \ ,
\ee
where the un-normalized correlator $\widetilde G(q)$ likewise only depends on the momentum through $q^2$. Fourier transforming this back to coordinate space gives
\be
 \delta \xi_+(x_1) = \int d^4 x_2 \, G(x_1-x_2) \delta \psi_-(x_2) 
\ee
where $G$ is the inverse Fourier transform of $\widetilde G$. The fact that $\widetilde G$ only depends on $q^2$ implies, among other things, that $G(x_1-x_2)=G(x_2-x_1)$. Therefore we can write
\be
 \frac{\delta \xi_+(x_1)}{\delta \psi_-(x_2)} = \frac{\delta \xi_+(x_2)}{\delta \psi_-(x_1)} = \int \frac{d^4q}{(2\pi)^4}\, \widetilde G(q)e^{iq_\mu (x_1^\mu-x_2^\mu)}
\ee
That is, we obtained the recipe to compute the functional derivatives in terms of $\widetilde G$ which can be directly extracted form the solutions to the fluctuation equations.

It is immediate that the additional term in the presence of the dilaton field in the VEV of the trace of the energy momentum tensor is
\be \label{eq:TmunutrV}
 \langle\theta^\mu_\mu(x)\rangle = -\frac{2\ell^3}{\kappa^2} \mathcal{N}(\Delta-2)(4-\Delta)X_+X_- -\frac{2\ell^3}{\kappa^2} (\Delta_g-2)(4-\Delta_g)\phi_s\phi_v 
\ee
where we reinstated the dilaton contribution.
Notice that it looks like setting here $\Delta_g=4$ would just eliminate the gluon contribution. This is, however, not that straightforward because this value is a special case. In particular, a constant source solution of $\phi$ will not represent a good leading RG flow end at $\phi=0$. Higher order corrections of the dilaton potential will modify the flow of source as we will se below in Appendix~\ref{app:RG_flow}.

Finally, we note that a similar computation yields the expression for the chiral condensate. That is, turning on $\chi_-$ instead of $\psi_-$, we find 
\bea
\langle\bar q q\rangle &=& i \frac{\delta}{\delta \chi_-(x)}\log Z_\mathrm{QCD}[\chi_-]\Big|_{\chi_-=0} = i \frac{\delta}{\delta \chi_-(x)}\exp(iS^{(1)}_\mathrm{ren})\Big|_{\chi_-=0} \nonumber \\
&=& -\frac{2(\Delta-2)\mathcal{N}\ell^3}{\kappa^2}X_+ \ .
\label{eq:chiralcfull}
\eea

\section{Holographic RG flow and the ward identity} \label{app:RG_flow}

In this appendix, we sketch how to obtain an expression of the correlator $\langle\theta^\mu_\mu(x)\rangle$ in terms of the  beta function and the gluon condensate. Following~\cite{Gursoy:2007cb}, we identify the 't Hooft coupling as $\lambda = e^\phi$ and the logarithmic of the energy scale as $A(r)\approx -\log r$. Therefore
\be \label{eq:betaDeltag}
 \beta(r) = -\frac{de^\phi}{d\log r} \approx - (4-\Delta_g)\phi_s r^{4-\Delta_g}e^{\phi_\uv} \ ,\qquad  (r \to 0) \ .
\ee
For the condensate we find using the dictionary and the expansion showed in equation \eqref{eq:Sbdry} that taking only the vev terms in the expansion for $\phi$ we get 
\be
 \langle  {\mathrm{Tr}} G_{\mu\nu}G^{\mu\nu} \rangle  \approx \frac{4\ell^3}{\kappa^2N_c}\left(\Delta_g-2\right)\phi_v \eps^{\Delta_g-4}e^{\phi_\uv}
\ee
Therefore the latter term in~\eqref{eq:TmunutrV} can be written as
\be  \label{eq:trG2Deltag}
 -\frac{2\ell^3}{\kappa^2} (\Delta_g-2)(4-\Delta_g)\phi_s\phi_v \approx \frac{N_c}{2\l^2} \beta(\eps)  \langle  {\mathrm{Tr}} G_{\mu\nu}G^{\mu\nu} \rangle 
\ee

Now we can combine the results~\eqref{eq:TmunutrV},~\eqref{eq:chiralcfull}, 
and~\eqref{eq:trG2Deltag} to the Ward identity
\be \label{eq:Wardfull}
  \langle\theta^\mu_\mu\rangle = (4-\Delta)m_q\langle\bar q q\rangle + \frac{N_c}{2\l^2} \beta(\eps)  \langle  {\mathrm{Tr}} G_{\mu\nu}G^{\mu\nu} \rangle \ .
\ee
Note that in the main text,~\eqref{eq:Ward}, we set $\Delta=3$.

\subsection{Different dilation potential examples}

One can choose a different potential for the dilaton from the one chosen in \eqref{eq:VphiUV} and see how the correlator \eqref{eq:TmunutrV} changes. A first natural example is
\be
\label{eq:Vphi4}
 V(\phi) = \frac{12}{\ell^2} +\lambda_g\phi^4
\ee
but since the computation is the same for all power low corrections, we can consider a more general case
\be
V(\phi) = \frac{12}{\ell^2} +\lambda_g\phi^n \ , \qquad (n>2) \ .
\ee 
Then the expansions for the dilaton and the metric are 
\bea \label{eq:nexpand}
 \phi(r) &\sim&\phi_s(-\log(r))^{\frac{1}{2-n}}\left(1+\mathcal{O}\left( \frac{1}{-\log(r)}\right)\right)+\\
 && \phi_v (-\log(r))^{\frac{n-1}{n-2}}r^{4}\left(1+\mathcal{O}\left(\frac{1}{-\log(r)}\right) \right) \nonumber \\
A(r) &\sim& \log\Big(\frac{\ell}{r}\Big)+\frac{\kappa_s }{(-\log (r))^{\frac{n}{n-2}}}\left(1+\mathcal{O}\left( \frac{1}{-\log(r)}\right)\right)-\kappa_v \phi_s \phi_v r^{4}\nonumber
\eea
\\
where the constants $\phi_s$,
$\kappa_s$, and $\kappa_v$ are given by
\bea
\phi_s= \bigg(\frac{8}{\ell^2\lambda_gn(n-2)}\bigg)^{\frac{1}{n-2}} \\
\kappa_s= -\frac{\ell^2\lambda_g}{24}\bigg(\frac{8 }{\ell^2\lambda_gn(n-2)}\bigg)^{\frac{n}{n-2}} \nonumber \\
\kappa_v=\frac{2}{15(n-2)}  \nonumber 
\eea
Note that the expansion of the metric also contains the terms arising from the backreaction of the $X$ field, which were written down in~\eqref{eq:AUVexp}, and suppressed here for brevity.

For the case of general $n>2$ the dilaton counterterm used in the previous case for $n=2$ changes. It reads
\be
 S_\mathrm{ct,\phi} = -\frac{\ell \lambda_g}{4\kappa^2\ell } \int d^4x\,\sqrt{-\det \gamma_b}\, \phi(r,x)^n\Big|_{r=\eps} \ .
\ee
Therefore the renormalized action becomes:
\bea \label{eq:action1b}
  S^{(1)}_\mathrm{ren} &=& \frac{1}{2\kappa^2}\frac{e^{4 A(r)}}{\ell}\int d^4x\,\Big[ - \left(\mathcal{N}(4-\Delta) X(r)^2+\frac{1}{4} \ell^2 \lambda_g \phi(r)^n+6 \ell e^{- A(r)} A'(r)+6\right) \nonumber\\
&&  \times\left(3\psi(r,x)+\tau(r,x)\right) +2 \mathcal{N}  \left(\ell e^{- A(r)} X'(r)-(4-\Delta) X(r)\right)\chi (r,x)\\ 
&& +2  \left(\ell e^{- A(r)} \phi'(r)-\frac{n}{8}\ell^2 \lambda_g \phi(r)^{n-1}\right)\varphi (r,x) \Big]\Big|_{r=\eps} \nonumber
\eea
After considering the boundary conditions from appendix  D.1 and inserting the ansatze for the fields the first-order action becomes
\bea
 S^{(1)}_\mathrm{ren} &\simeq& - \frac{2\ell^3}{\kappa^2\eps^4}\left(\mathcal{N}(4-\Delta) X(\eps)^2+6 \ell e^{-A(\eps)} A'(\eps)+6+\frac{\ell^2}{4}\lambda_g\phi^n(\epsilon)\right) \int d^4x\ \psi_-\nonumber \\ 
 &=&  -\frac{2\ell^3}{\kappa^2} \Big(\mathcal{N}(2-\Delta)(4-\Delta)X_+X_- + \frac{2\phi_s \phi_v}{(n-2)}\Big)\int d^4x\ \psi_-
 \eea
Thus the correlator reads
\bea 
 \langle\theta^\mu_\mu(x)\rangle &=& i \frac{\delta}{\delta \psi_-(x)}\log Z_\mathrm{QCD}[\psi_-]\Big|_{\psi_-=0}=- \frac{\delta}{\delta \psi_-(x)} S^{(1)}_\mathrm{ren}[\psi_-]\Big|_{\psi_-=0}  \nonumber\\
 &&=\frac{2\ell^3}{\kappa^2}\Big(\mathcal{N}(2-\Delta)(4-\Delta)X_+X_- + \frac{2\phi_s \phi_v}{(n-2)}\Big) \nonumber\\
\eea

Now we can express this in terms of the beta function and the condensate. Using the expansion for $\phi$ in \eqref{eq:nexpand}, the beta function is given by
\bea
\beta(r) \approx -\frac{\phi_s e^{\phi_\uv}}{(2-n)\left(-\log(r)  \right)^{\frac{n-1}{n-2}}}  \ ,\qquad  (r \to 0) \ .
\eea
Then again using only  the terms of the expansion of $\phi$ proportional to $r^4$ we find  the  condensate:
\bea
 \langle  {\mathrm{Tr}} G_{\mu\nu}G^{\mu\nu} \rangle  \approx \frac{8\ell^3 \phi_v}{\kappa^2N_c}\left[\left(-\log(\epsilon) \right)^{\frac{n-1}{n-2}}-\frac{n^2 \ell^2}{32}\lambda_g \phi^{n-1}_s \phi_v  \right] e^{\phi_\uv}
 \eea
Therefore we find that 
\be
\frac{N_c}{4\l^2} \beta(\eps)  \langle  {\mathrm{Tr}} G_{\mu\nu}G^{\mu\nu} \rangle \approx \frac{2 \ell^3 \phi_s \phi_v}{\kappa^2 (n-2)}
\ee
Collecting the results, and also recalling~\eqref{eq:chiralcfull}, we find the expected Ward identity~\eqref{eq:Wardfull}.

We can also  explore the potential  which has the UV asymptotics used in the context of the IHQCD model~\cite{Gursoy:2007cb,Gursoy:2007er}
\be 
V(\phi) =  \frac{12}{\ell^2}(1 +v_1e^{\frac{\sqrt{3}}{2}\phi})
\ee 
where we only included the leading terms at small $\phi$ and the factor of $\sqrt{3}/2$ appears because of our normalization convention for the kinetic term of the dilaton field.
In IHQCD, the UV fixed point is at $\phi =- \infty
$ and the UV running of the dilaton is driven by corrections to the potential suppressed by powers of $e^{\phi}$. In this case the expansion of the dilaton is the following 
\bea
 \phi(r) &=&\Big(-\frac{2}{\sqrt{3}}\log\big(-\phi_s\log(r)\big)\Big) + \phi_v (-\log(r))r^{4}\left(1+\mathcal{O}\left( \frac{1}{-\log(r)}\right)\right)\nonumber \\
A(r) &=& \log\Big(\frac{\ell}{r}\Big)+\frac{\kappa_s }{-\log (r)}-\kappa_v \phi_s \phi_v r^{4}
\eea 
where the constants $\phi_s$,
$\kappa_s$, $\kappa_v$ are given by
\bea
\phi_s= \frac{9 v_1}{8} \\
\kappa_s= -\frac{4}{9}\nonumber \\ 
\kappa_v= -\frac{32}{135 \sqrt{3} v1} \nonumber 
\eea

For the case of IHQCD we find that the counterterm for the dilaton reads
\be
 S_\mathrm{ct,\phi} = -\frac{3 v_1 }{2 \kappa^2 \ell} \int d^4x\,\sqrt{-\det \gamma_b}\, e^{\frac{\sqrt{3}}{2}\phi(r,x)}\Big|_{r=\eps} \ .
\ee
Therefore the action becomes:
\bea \label{eq:actionIHQCD}
  S^{(1)}_\mathrm{ren} &=& \frac{1}{2\kappa^2}\frac{e^{4 A(r)}}{\ell}\int d^4x\,\Big[ - \left(\mathcal{N}(4-\Delta) X(r)^2+6 \ell e^{- A(r)} A'(r)+6+3 v_1 e^{\frac{\sqrt{3}}{2}\phi(r)}\right) \nonumber\\
&&  \times\left(3\psi(r,x)+\tau(r,x)\right) +2 \mathcal{N}  \left(\ell e^{- A(r)} X'(r)-(4-\Delta) X(r)\right)\chi (r,x)\\ 
&& +2  \left(\ell e^{- A(r)} \phi'(r)-\frac{3\sqrt{3} v_1}{4}e^{\frac{\sqrt{3}}{2}\phi(r)}\right)\varphi (r,x) \Big]\Big|_{r=\eps} \nonumber
\eea

After consider the boundary conditions from appendix  D.1 and inserting the Ansatz for the fields the first-order action becomes 
\bea
 S^{(1)}_\mathrm{ren} &\simeq& - \frac{2\ell^3}{\kappa^2\eps^4}\left(\mathcal{N}(4-\Delta) X(\eps)^2+3 v_1e^{\frac{\sqrt{3}}{2}\phi(r)}+6 \ell e^{-A(\eps)} A'(\eps)+6\right) \int d^4x\ \psi_-\nonumber \\ 
 &=&  -\frac{2\ell^3}{\kappa^2} \Big(\mathcal{N}(2-\Delta)(4-\Delta)X_+X_- - \frac{4\phi_v}{\sqrt{3}}\Big)\int d^4x\ \psi_-
 \eea
Thus the correlator reads 
\bea \label{eq:Tmunuforphin}
 \langle\theta^\mu_\mu(x)\rangle &=& i \frac{\delta}{\delta \psi_-(x)}\log Z_\mathrm{QCD}[\psi_-]\Big|_{\psi_-=0}=- \frac{\delta}{\delta \psi_-(x)} S^{(1)}_\mathrm{ren}[\psi_-]\Big|_{\psi_-=0}  \nonumber\\
 &&=\frac{2\ell^3}{\kappa^2} \Big(2\mathcal{N}(4-\Delta)X_+X_- - \frac{4\phi_v}{\sqrt{3}}\Big) \nonumber\\
\eea

For this case the beta function is given by:
\bea
\beta(r) \approx\frac{2}{\sqrt{3}}\left( \frac{1}{-\log(r)}\right) e^{\phi(r)},\qquad  (r \to 0)
\eea
and the condensate is given by:
\bea
\langle  {\mathrm{Tr}} G_{\mu\nu}G^{\mu\nu} \rangle  \approx \frac{8\ell^3 \phi_v}{\kappa^2N_c}\left[\log(r)+\phi_{v2}-\kappa_s\right] e^{\phi(r)}
 \eea
Therefore we find that 
\be
\frac{N_c}{4\l^2} \beta(\eps)  \langle  {\mathrm{Tr}} G_{\mu\nu}G^{\mu\nu} \rangle \approx -\frac{4 \ell^3 \phi_v}{\sqrt{3}\kappa^2}
\ee
Collecting the results, we again verify the Ward identity~\eqref{eq:Wardfull}.

\section{On the generality of the IR boundary conditions}\label{app:bcs}

In this Appendix, we discuss the IR boundary conditions for the field $X$. We focus on the background solution in here, but similar comments apply to the fluctuations, as we argue at the end of the Appendix. 

In the main text, we used generic IR boundary conditions (see Eqs.~\eqref{eq:IRbcgen} and~\eqref{eq:IRbcgenqq}). The way we define the boundary conditions, in particular in the model computation of Sec.~\ref{sec:toymodel}, may appear dubious because the definition~\eqref{eq:IRbcgen} on the one hand assumes linearity of the equation of motion of $X$ but on the other hand in the IR regime $r \sim r_\ir$ the field $X$ will grow large so that non-linearities and backreaction to the metric become important. In the analytic model per se there is no problem, as non-linearities and backreaction are absent by definition, but we argue that the model also represents setups where these features are included. Here we show why this is the case.

We begin by considering a generic action for $X$ which we do not specify explicitly, but we assume that it is backreacted to the gravity sector and agrees with~\eqref{eq:Xaction} when $X \ll 1$. We denote the backreacted regular solution of the potentially non-linear action by $X_\mathrm{nl}$. For concrete examples (such as the model in~\cite{Jarvinen:2011qe}) such a solution can be found numerically. Without picking an explicit model we cannot specify what is meant by regularity, but for the model to be consistent and able to describe walking geometries, there needs to be a way to pick a single solution such that $X$ becomes small when evolved towards the boundary, while the geometry approaches the IR fixed point in the walking regime, $r_\uv \ll r \ll r_\ir$. That is, the background takes the form of~\eqref{eq:bgwalking} while the flavor field obeys~\eqref{eq:XIRgen}: 
\be \label{eq:Xnl}
  X_\mathrm{nl}(r) = X_\ir^\mathrm{nl} \left(\frac{r}{r_\ir}\right)^2 \sin\left(\nu \log \frac{r}{r_\ir}-k_\ir^\mathrm{nl}\right) \ , \qquad (r_\uv \ll r \ll r_\ir) \ .
\ee
For these equations to make sense we also need to define the UV and IR scales. We can use the definitions from Sec.~\ref{sec:RGflow}, i.e., Eqs.~\eqref{eq:rUVdef} and~\eqref{eq:rIRdef}. The regular solution must be unique, so that $X_\ir^\mathrm{nl}$ and $k_\ir^\mathrm{nl}$ are also uniquely defined through~\eqref{eq:Xnl}. 

We then consider a general solution of $X_\mathrm{lin}$ to the linearized action and in the absence of backreaction. This solution obeys\footnote{Note that in the absence of backreaction, the background stays at the IR fixed point for all $r\gg r_\uv$.}
\be \label{eq:Xlin}
 X_\mathrm{lin} =X_\ir^\mathrm{lin} \left(\frac{r}{r_\ir}\right)^2 \sin\left(\nu \log \frac{r}{r_\ir}-k_\ir^\mathrm{lin}\right) \ , \qquad (r \gg r_\uv)
\ee
where we can take the UV and IR scales to be the same as for the non-linear solution. Now we can set the IR regularity condition for the linear solution to be 
\be \label{eq:IRgencond}
 X_\ir^\mathrm{lin} = X_\ir^\mathrm{nl} \ , \qquad k_\ir^\mathrm{lin} =k_\ir^\mathrm{nl} \ ,
\ee
which is well defined in the limit $r_\ir/r_\uv \to \infty$. 

Now the interpretation for the condition~\eqref{eq:IRbcgen} is clear: it should be required for the linearized solution $X_\mathrm{lin}$, which obeys the simple form of~\eqref{eq:Xlin} even for $r \sim r_\ir$, rather than $X_\mathrm{nl}$. This IR condition is equivalent to  the condition for the phase in~\eqref{eq:IRgencond} if we set 
\be
 \frac{A}{B} = \frac{\nu}{\tan k_\ir^\mathrm{nl}} -2 \ .
\ee
This condition does not determine the normalization, $X_\ir$, but in the absence of backreaction and non-linearities this parameter scales out from the action so it affects the analysis only trivially. Note, however, that it is a number which is independent of the UV scale $r_\uv$, and is also determined by the definition of $r_\ir$.

One can follow similar steps to show that general IR regularity conditions for the fluctuation $\xi$ can be captured by our conditions~\eqref{IRbc}. The fluctuation equations are always linear, but the general regular solution should take into account the coupling between the wave functions $\xi$ and $\zeta$ in~\eqref{eq:xieq} and~\eqref{eq:zetaeq} which becomes sizable for $r \sim r_\ir$. This coupling can be treated in the same way as the non-linearities for the background solution. There is however an important complication: the coefficients (for example the counterpart of $X_\ir^\mathrm{nl})$ will depend on the mass or frequency of the fluctuations $\omega$. However if the mass is small, $\omega \ll 1/r_\ir$, it can be neglected in the analysis. That is, the coefficients $\mathcal{A}$, $\mathcal{B}$ in~\eqref{IRbc} can be interpreted to be fixed numbers only for light modes.

\bibliography{hdilaton}
\bibliographystyle{utphys}

\end{document}